\documentclass[useAMS,usenatbib]{mnras}
\usepackage{amsmath}
\usepackage{amssymb}
\usepackage{graphicx}
\usepackage{array}
\usepackage{tabularx}
\usepackage{float}

\def \farcs{\hbox{$.\!\!^{\prime\prime}$}}

\title[CCCP and MENeaCS]
  {CCCP and MENeaCS: (updated) weak-lensing masses for 100 galaxy clusters}
\author[Ricardo Herbonnet et al.]
{\parbox[t]{\textwidth}{\vspace{-0.7cm}
\begin{flushleft}
  Ricardo~Herbonnet$^{1,2}$\thanks{Email: ricardo.herbonnet@stonybrook.edu}, 
  Crist\'{o}bal~Sif\'{o}n$^{3,2}$, 
  Henk~Hoekstra$^2$, 
  Yannick~Bah\'{e}$^2$, 
  Remco~F.~J.~van~der~Burg$^{4}$, 
  Jean-Baptiste~Melin$^{5}$, 
  Anja~von~der~Linden$^{1}$,
  David~Sand$^{6}$,
  Scott~Kay$^{7}$,
  David~Barnes$^{8}$
  \end{flushleft}
  }
  \\
  $^1$Department of Physics and Astronomy, Stony Brook University, Stony Brook, NY 11794, USA\\
  $^2$Leiden Observatory, Leiden University, PO Box 9513, 2300 RA, Leiden, The 
Netherlands\\
  $^3$Instituto de F\'isica, Pontificia Universidad Cat\'olica de Valpara\'iso, Casilla 4059, Valpara\'iso, Chile\\
  $^4$European Southern Observatory, Karl-Schwarzschild-Str. 2, 85748, Garching, Germany\\
  $^5$IRFU, CEA, Universit{\'e} Paris-Saclay, F-91191 Gif-sur-Yvette, France\\
  $^6$Department of Astronomy/Steward Observatory, 933 North Cherry Avenue, Rm. N204, Tucson, AZ 85721-0065, USA\\
  $^7$Jodrell Bank Centre for Astrophysics, Department of Physics and Astronomy, The University of Manchester,\\ Manchester M13 9PL, UK\\
  $^8$Department of Physics, Kavli Institute for Astrophysics and Space Research, Massachusetts Institute of Technology,\\ Cambridge, MA 02139, USA​
}
\date{9 December 2019}

\pubyear{2019}

\pagerange{\pageref{firstpage}--\pageref{lastpage}} 

\def\LaTeX{L\kern-.36em\raise.3ex\hbox{a}\kern-.15em
    T\kern-.1667em\lower.7ex\hbox{E}\kern-.125emX}

\newcommand{\mn}{MENeaCS}

\newcommand{\pz}{photo-$z$}

\defcitealias{hoekstra15}{H15}

\begin{document}

\label{firstpage}

\maketitle

\begin{abstract}
Large area surveys have detected significant samples of galaxy clusters that can be used to constrain cosmological parameters, provided that the masses of the clusters are measured robustly. To improve the calibration of cluster masses using weak gravitational lensing we present new results for 48 clusters at $0.05<z<0.15$, observed as part of the Multi Epoch Nearby Cluster Survey (MENeaCS), and reevaluate the mass estimates for 52 clusters from the Canadian Cluster Comparison Project (CCCP).
Updated high-fidelity photometric redshift catalogues of reference deep fields are used in combination with advances in shape measurements and state-of-the-art cluster simulations, yielding an average systematic uncertainty in the lensing signal below 5\%, similar to the statistical uncertainty for our cluster sample. We derive a scaling relation with \textit{Planck} measurements for the full sample and find a bias in the \textit{Planck} masses of $1-b=0.84 \pm 0.04$.  We find no statistically significant trend of the mass bias with redshift or cluster mass, but find that different selections could change the bias by up to 1.5$\sigma$. We find a gas fraction of $0.139 \pm 0.014$ for 8 relaxed clusters in our sample, which can also be used to infer cosmological parameters. 
\end{abstract}


\section{Introduction}
\label{intro}

The growth rate of massive haloes is sensitive to cosmology as the gravitational build-up of overdensities in the initial density distribution is counteracted by the expansion of the Universe. Numerical simulations can predict the abundance of massive haloes for varying cosmologies and linking these to such objects in the real Universe allows for cosmological tests (see \citealt{allen11} for a general review). Although the bulk of the mass in these structures is in the form of dark matter, they are observable across the electro-magnetic spectrum because they contain large amounts of baryons that manifest their presence in various ways, such as clusters of galaxies and hot gas.
Studies of the number of clusters as a function of mass and redshift (cluster mass function) have put tight constraints on the energy density of matter $\Omega_m$ and normalisation of the matter power spectrum $\sigma_8$ \citep[e.g.][]{borgani01, vikhlinin09b,rozo10}, and the redshift evolution of the mass function can constrain the abundance and the equation of state of dark energy, as well as the number of neutrino species \citep[e.g.][]{mantz10b, mantz15, planckclustercosmo16, dehaan16, bocquet19}. 

The determination of the cluster mass function requires a large sample of clusters with a well-defined selection function and accurate mass estimates of those clusters. The number of observed clusters is steadily increasing thanks to optical searches for overdensities of (red) galaxies \citep[e.g.][]{gladders05, rykoff16}, and X-ray surveys looking for diffuse hot intracluster gas \citep[e.g.][]{ebeling98,ebeling01,bohringer04, vikhlinin09a}. In recent years millimeter wavelength searches for the signatures of the Sunyaev-Zeldovich effects \mbox{\citep[SZ effect]{sz72}} in the cosmic microwave background (CMB) have added greatly to the number of detected clusters  \citep{planckSZclusters16,hilton18, bleem19}. CMB photons are present at all observable redshifts and the SZ signal scales linearly with gas density, making it observable even for high redshift clusters with relatively low gas density, promising many thousands of newly detected clusters in the near future.

Another requirement for robust estimates of cosmological parameters is a well calibrated relation between survey observable and mass\footnote{Because of degeneracies between cosmological and astrophyical parameters in the estimation, the masses and scaling relation should be inferred simultaneously with cosmological parameters \citep[e.g.][]{mantz10}}. In fact, the lack of a reliable scaling relation is the main limitation for the full exploitation of the already available CMB cluster catalogues. The total mass of clusters can be computed using kinematics of cluster members under the assumption of dynamical equilibrium \citep[e.g.][]{sifon16, amodeo17, armitage18} or using caustics \citep{rines16}. However, these estimates generally have large biases and/or large scatter \citep{old18}. X-ray measurements can be connected to mass, but this is usually done under the assumption of hydrostatic equilibrium. This assumption can lead to masses underestimated by $\sim$10-35\% depending on the dynamical state of the cluster \citep[e.g.][]{ henson17, barnes17}.

Fortunately, a galaxy cluster acts as a lens because its gravitational potential distorts the surrounding space-time, which deflects photons from their straight line trajectories. This phenomenon, known as gravitational lensing, introduces a coherent distortion (shear) in the observed shape of background galaxies, which scales with cluster mass. Most galaxies are only slightly sheared by the cluster and the statistical inference of the shear signal from a sample of background galaxies is known as weak gravitational lensing. Weak-lensing thus provides the total mass of a cluster without strict assumptions on the dynamical state of the cluster. Simulations show that lensing mass estimates are nearly unbiased, so other mass proxies can be calibrated against it. However, the triaxial distribution of mass introduces a scatter of $\sim$10-30\% in lensing masses for individual cluster \citep{becker11, rasia12, bahe12, henson17,herbonnet19}. There is also a large statistical uncertainty in the shear, which is obtained by averaging of background galaxy shapes. Moreover, uncorrelated large scale structure introduces extra scatter in the mass estimates \citep{hoekstra01, hoekstra11}. For a large sample of clusters these should average out, so reliable scaling relations can only be produced for large samples of clusters.
This has been the subject of numerous studies \citep[e.g.][]{okabe13, umetsu14, vonderlinden14, hoekstra15, okabe16, schrabback18,mcclintock18, miyatake19, bellagamba19,nagarajan19}.


Weak-lensing experiments measure the shear by averaging the shapes of galaxies behind the clusters, and combine these with distance estimates for the background galaxies in order to reconstruct the mass profile. The background galaxies are predominantly faint objects, so their distances are computed using photometric redshifts. Systematics are thus introduced by biased measurements of the galaxy shapes and of the galaxy redshifts, a false classification of objects as background galaxies, and incorrect assumptions of the mass profile of the cluster. 
\citet[ hereafter \citetalias{hoekstra15}]{hoekstra15} performed a thorough analysis of most of these systematics for the Canadian Cluster Comparison Project (CCCP), finding agreement with the independent, equally thoroughly calibrated, pipeline of the Weighing the Giants (WtG) project \citep{vonderlinden14, applegate14} for clusters observed in both surveys. 

In this work, we build on the work of \citetalias{hoekstra15} by studying another sample of clusters, which was also observed with the Canada-France-Hawaii Telescope (CFHT), as was CCCP, and analyse it with the same pipeline. The Multi Epoch Nearby Cluster Survey (\mn) provides excellent quality optical imaging data in the $g$ and $r$-band for a sample of 58 X-ray selected clusters at $0.05<z<0.15$. \mn \ presents a significant collection of clusters allowing for a precise determination of the average cluster mass. The low redshift range, and hence small volume, in combination with the steepness of the halo mass function at cluster scales, means that \mn \ clusters are on average less massive than CCCP clusters, thereby also extending the mass range for the scaling relation analysis. However, the trade-off for our large sample size is the lack of colour information required to estimate photometric redshifts (\pz's) for all observed galaxies. Fortunately, new deep high fidelity \pz \ catalogues of reference fields have become available to address this issue. Therefore, in addition to presenting the cluster masses for \mn, we will also update the mass estimates for CCCP clusters in this work.

The \mn \ observations are briefly described in Section~\ref{sec:data}, where we also present details of the pipeline used to determine galaxy shapes. 
In Section~\ref{sec:photoz} we determine a distribution of redshifts for the background galaxy population. Without reliable redshift information for individual objects, galaxies cannot be separated into a population associated to the cluster and a population of gravitationally lensed background galaxies. This is addressed in Section~\ref{sec:contamination}. Section~\ref{sec:masses} describes the determination of the cluster masses, which are compared to other mass estimates in Section~\ref{sec:hsebias} and we conclude in Section~\ref{sec:conclusions}. Throughout the paper we assume a flat $\Lambda$ cold dark matter cosmology where $H_0$=70 km/s/Mpc and the current energy densities of matter and dark energy are $\Omega_m(z=0)=0.3$ and $\Omega_\Lambda(z=0)=0.7$, respectively.

\section{Data and shear analysis}
\label{sec:data}

\subsection{\mn \ data}

The Multi Epoch Nearby Cluster Survey (\mn) is a deep, wide-field imaging survey of a sample of X-ray selected clusters with $0.05<z<0.15$. The data were obtained with two main science objectives in mind. The first, the study of the dark matter halos of cluster galaxies using weak gravitational lensing, defined the required total integration time and image quality, as well as the redshift range. The results of this analysis are presented in \citet{sifon18b, sifon18}. Taking advantage of the queue scheduling of CFHT observations, however, the observations were spread over a two-year period, which enabled a unique survey to study the rate of supernovae in clusters \citep{sand12, graham12, graham15}, including intra-cluster supernovae \citep{sand11}. To do so, typically two 120s exposures in the $g$ and $r$-band were obtained for each epoch (which are a lunation apart).  
The full sample comprises the 58 most X-ray luminous clusters that are observable with the CFHT. A detailed description of the survey is presented in \citet{sand11, sand12}. 

In this paper we use the $r$-band data to determine the \mn \ cluster masses using weak gravitational lensing. The individual exposures are pre-processed using the {\tt Elixir} pipeline \citep{magnier04}, and we refine the astrometry using {\tt Scamp} \citep{scamp}. Although the CFHT observations were typically obtained when the seeing was below $1''$, some exposures suffer from a larger PSF. As this is detrimental for accurate shape measurements, these exposures were excluded when co-adding the data. For each cluster the 20 frames with the best image quality were selected and combined into a single deep coadded image using {\tt Swarp} \citep{bertin10}. However, if additional frames had a seeing full width at half maximum less than $0.80''$ they were added to the stack. The minimal depth of each coadded image is therefore 40 minutes of exposure time. 
The magnitudes we use are corrected for Galactic extinction using the \cite{schlafly11} recalibration of the \cite{schlegel98} infrared-based dust map. 
For the analysis presented here, we exclude 9 clusters based on their $r$-band Galactic dust extinction $A_r$. 
The threshold value $A_r<0.2$ was chosen to reflect the range in which we can reliably correct for contamination (see Section~\ref{sec:contamination} and Appendix~\ref{app:blanks}).
Finally, the cluster Abell~763 contains no significant overdensity of galaxies, nor is it part of the \textit{Planck} cluster catalogue, and was removed from the sample. Table~\ref{tbl:redshifts}\footnote{To improve readability we show all large tables in the appendix.} lists for all selected clusters their properties and for \mn \ clusters the characteristics of the observations. The coordinates of the brightest cluster galaxy (BCG) are taken as the centre of the cluster. The BCGs were selected based on a visual inspection of the data \citep{bildfell08}.

\subsection{Source selection}

Objects were detected in the coadded images  using the pipeline described in \citet{hoekstra12}. To measure the weak-lensing signal around the clusters we select objects with an $r$-band magnitude $20\leq m_r \leq 24.5$. Following \citetalias{hoekstra15} an upper limit of 5 pixels on the galaxy half-light radius is imposed to help remove spurious detections, such as blended objects, from the object catalogue. A lower limit on the size is set by the size of the PSF, which removes stars and small galaxies that have highly biased shapes. 

Galaxy magnitudes are corrected for background light by subdividing pixels in an annulus between 16 and 32 pixels into four quadrants and fitting the quadrants with a plane to allow for spatial variation of the background. We found that bright neighbouring objects affect this local background subtraction, which in turn affects the shape measurement. When we examined the performance of the algorithm near bright cluster members in image simulations for the purpose of studying the lensing signal around such galaxies \citep{sifon18}, there were cases where $m_{\rm det}$, the apparent magnitude as measured by the detection algorithm differed from  $m_{\rm shape}$, the magnitude measured by the shape measurement algorithm. This change in magnitude was introduced by the background subtraction algorithm. No background light was present in the simulations and instead the local background subtraction was affected by the light of nearby bright cluster galaxies. An empirically derived relation based on  $\Delta m=m_{\rm det}-m_{\rm shape}$ of 
\begin{equation}
\Delta m > 49.0 -7.0 \hspace{1pt} m_{\rm shape} +0.3 \hspace{1pt} m_{\rm shape}^2 -0.005 \hspace{1pt} m_{\rm shape}^3
\label{eq:delmag}
\end{equation} 
efficiently identified these problematic objects in the simulations. We therefore apply this cut to the data, which removes a few percent of the detected objects. 

\subsection{Shear measurement}

The galaxy polarisations and polarisabilities are measured from the mosaics using the shape measurement algorithm detailed in \citetalias{hoekstra15}, which is based on the moment-based method of \cite{kaiser95}. The polarisation $\chi$  is a measure of the galaxy ellipticity and is determined using a weight function to reduce the effect of noise, which introduces a bias in the final shear estimate. The shear polarisability $P^\gamma$ corrects the polarisation for the use of the weight function and for the effect of the PSF. Galaxies are assigned a lensing weight
\begin{equation}
 w = \left[ \langle \epsilon_{\mathrm{int}}^2 \rangle + \left( \dfrac{\sigma_\chi}{P^\gamma} \right)^2 \right]^{-1} ,
 \label{eq:lensweight}
\end{equation}
where $\langle \epsilon_{\mathrm{int}}^2 \rangle=0.25^2$ is the dispersion in the distribution of intrinsic ellipticities and $\sigma_\chi$ is an estimate of the uncertainty in the measured value of $\chi$ due to noise in the image \citep{hoekstra00}. The shear for an ensemble of galaxies is computed as the weighted average of the corrected polarisations 
\begin{equation}
 g_i = \dfrac{\sum\limits_n w_n \chi_{i,n}/P^\gamma_n}{\sum\limits_n w_n} ,
 \label{eq:redshear}
\end{equation}
where the index $i$ indicates the two Cartesian components of the shear and the sum runs over all galaxies in the sample. 
In practice, we measure the reduced shear $g_i=\gamma_i/(1-\kappa_i)$, where $\gamma$ is the true shear of the object and the convergence $\kappa$ is a measure of the magnification and change in size of an object due to gravitational lensing \citep{bartelmann01}. The reduced shear $g$ therefore deviates from the true shear $\gamma$. However, for most radii of interest $\kappa$ is very small and the difference between $g$ and $\gamma$ is negligible, although we take it into account in our analysis. Henceforth, we refer to the reduced shear $g$ as the shear. We decompose the shear into a cross and tangential component relative to the lens, where the tangential shear $g_\mathrm{t}$ can be related to the projected mass of the lens and the cross shear can be used to find systematic errors \citep{schneider03}.

\citetalias{hoekstra15} used extensive image simulations to quantify the multiplicative bias that arises from noise in the data and the imperfect correction for blurring by the PSF. The MENeaCS data are similar in terms of depth and image quality compared to the observations of the CCCP that were analysed in \citetalias{hoekstra15}; therefore we use the same correction scheme. The correction is a function of the signal-to-noise ratio (SNR) and the measured size of the galaxies. A potentially important difference with the CCCP analysis is that the individual exposures are offset from one another. This could lead to a complicated PSF pattern in the combined images. However, tests on the CCCP data indicate that this results in a negligible change in the mass estimates. Moreover, the large number of exposures, combined with the smooth PSF pattern results in a smooth PSF when measured from the mosaics. We applied the selection of Equation \ref{eq:delmag} to the image simulations studied in \citetalias{hoekstra15} and found that the shear biases were unchanged. Consequently, we use the same parameters as they used to correct for the biases in the method. \citetalias{hoekstra15} estimated that the systematic uncertainties in the cluster masses caused by the shape measurements is less than $2\%$, which is also adequate for the results presented here. The image simulations did not have input shears larger than 0.07, so that the calibration is not reliable for larger shears. Therefore we restrict our analysis to data beyond 0.5 Mpc from the cluster centre, where shears are small enough to be reliably calibrated.

\section{Photometric source redshift distribution}
\label{sec:photoz}

Gravitational lensing is a geometric phenomenon and the amplitude of the effect depends on the distances involved. This dependency is parametrised by the critical surface density
\begin{equation}
 \Sigma_\mathrm{crit} =\frac{c^2}{4 \pi G} \frac{1}{D_\mathrm{ol} \hspace{1pt} \beta }, 
 \label{eq:sigcrit}
\end{equation}
where the lensing efficiency $\beta = \mathrm{max(}0, D_\mathrm{ls}/D_\mathrm{os})$ contains the redshift information about the background galaxy (termed the `source'). The angular diameter distances $D_\mathrm{os}, D_\mathrm{ls}, D_\mathrm{ol}$ are measured between observer `o', lens `l' and/or source `s'. The definition of $\beta$ is such that objects in front of the cluster, which are not gravitationally sheared, do not contribute to the measured signal. For an increasing source redshift the lensing efficiency $\beta$ rises sharply when the source is behind the lens, but it flattens off when source and lens are far apart.

We lack photometric information to compute redshifts for individual objects in our catalogue and hence we cannot determine the critical surface density for each source lens pair. However, as the galaxies are averaged to obtain a shear estimate, we can use an average lensing efficiency $\langle \beta \rangle$ to compute the critical surface density for the full source population. This assumption introduces a bias in our shear estimates which can be approximately corrected for by multiplying our shear estimates by
\begin{equation}
 \frac{\langle\hspace{3pt}  g(\beta)\hspace{3pt}  \rangle}{\langle \hspace{3pt} g(\langle \beta \rangle) \hspace{3pt} \rangle}  \approx 1+\left(\dfrac{\langle \beta^2 \rangle}{\langle \beta \rangle^2} -1\right) \kappa 
 \label{eq:photzbias}
\end{equation}
\citep[Equation 7 in][]{hoekstra00}. The numerator $\langle g(\beta) \rangle$ is the average shear using a redshift for each source and the denominator $\langle g(\langle \beta \rangle) \rangle$ is the average shear using an average lensing efficiency for the whole population of sources $\langle \beta \rangle$.  The width of the distribution of the lensing efficiency $\langle \beta^2 \rangle$ corrects the shear for the use of a single value of $\langle \beta \rangle$. For our local clusters most sources are so distant that there is little variation in the value of $\beta$. Indeed, we find that the ratio $\langle \beta^2 \rangle / \langle \beta \rangle^2 \approx 1$ for most clusters and so this correction is very small for our analysis\footnote{\citet{applegate14} used a slightly different expression for this correction, but given the small impact of Equation~\ref{eq:photzbias} changing this expression should not alter any of our results.}.

A reference sample of field galaxies can serve as a proxy for the source population in our observations in order to compute $\langle \beta \rangle$. For this we use the COSMOS field which has received dedicated deep photometric and spectroscopic coverage so that reliable redshift estimates are available. In our analysis we use the latest COSMOS2015 catalogue of \citet{laigle16} containing \pz's based on over 30 different filters. This catalogue has two important benefits for our analysis. First, near-infrared data from the UltraVISTA DR2 are included, so that the Lyman and Balmer/4000 \AA \ breaks can be distinguished. The additional knowledge on these features helps to address the degeneracy between low and high redshift galaxies. Second, the catalogue also includes the CFHT \textit{r} filter, so that we can easily match it to our data. Although the objects in the COSMOS2015 catalogue were not selected based on their $r$-band magnitude, we find that the catalogue is nearly complete down to $m_r \approx 25$,  sufficient to cover the full magnitude range $20 \leq m_r \leq 25$ for all our clusters. From comparisons to spectroscopic data \citet{laigle16} found that their redshift estimates are accurate to better than a percent, and 2\% for high redshift galaxies, which is sufficient for this study. We select galaxies from the matched catalogue using the \texttt{TYPE} parameter, which classifies objects as either stars or galaxies. 

The COSMOS2015 catalogue is not representative of our lensing catalogues, as the latter are subject to various cuts (Section~\ref{sec:data}). \citet{gruen17} have shown that these selection effects can introduce a bias in the mass estimates. To account for this, we ran our lensing pipeline on $r$-band observations of the CFHT Legacy Survey (CFHTLS) D2 field which covers $\sim$1 square degree of the COSMOS field and matched the lensing catalogue to the COSMOS2015 catalogue. This enabled us to match the cuts on the lensing data to the redshift distribution.
We found that applying the cuts introduces a difference in the lensing efficiency of only $\sim$0.5\% of $\langle \beta \rangle$ for all clusters. We use the matched catalogue for our \pz \ analysis, but note that the addition of the cuts does not significantly impact our results, nor the results of \citetalias{hoekstra15}.

Even after applying the same cuts there may still be differences in the distributions of lensing weights, used in the shear estimation, in our data and in the COSMOS field due to different seeing conditions. Consequently, directly using the \pz \ distribution from COSMOS for our lensing analysis can lead to biases. Therefore we customise our COSMOS galaxy population according to the galaxy population in each cluster observation, similar to the procedure in \citetalias{hoekstra15}.
To do this the \pz \ catalogue is divided into magnitude bins. For each magnitude bin $i$ we compute the sum of the lensing weights of the COSMOS galaxies in that bin and the mean lensing efficiency $\langle \beta \rangle^\mathrm{COSMOS}_i$. Then for the same magnitude bins we compute the sum of the lensing weights in the cluster data. The lensing weights are used as a reweighting factor $R_i$ for the COSMOS magnitude distribution to match the distribution observed for the cluster. The final estimate average lensing efficiency for a cluster is  
 \begin{align}
 & \langle \beta \rangle = \sum_i \left( \langle \beta \rangle^\mathrm{COSMOS}_\mathrm{i} R_i \right) / \sum_i  \left( R_i \right) \nonumber \\
 & R_i =  \frac{\sum_j w^\mathrm{cluster}_{j,i} }{ \sum_j w^\mathrm{COSMOS}_{j,i}} ,
 \end{align}
where the subscript $i$ designates a magnitude bin and $j$ the objects falling into that bin.
For each cluster the value of $\langle \beta \rangle$ is listed in Table~\ref{tbl:redshifts}. We use $\langle \beta \rangle$ to compute the average critical surface density with which we estimate cluster masses. In order to apply Equation~\ref{eq:photzbias} we also require  $\langle \beta^2 \rangle$, which is calculated the same way and listed in Table~\ref{tbl:redshifts}. The higher values of $\beta$ for the CCCP clusters at similar redshifts as \mn \ clusters are due to the different magnitude range 22-25, compared to 20-24.5 for \mn. Magnification by the cluster can change the distribution of magnitudes and redshifts of background galaxies compared to a reference field. We checked the effect of changing the magnitude ranges by 0.02 magnitudes, as an estimate of the effect of magnification by the cluster and found that this has only a small effect on $\langle \beta \rangle$.

The redshift distribution in our catalogue is based on 1 square degree of the COSMOS field and might not be representative for all source populations in our observations. This cosmic variance introduces an uncertainty in the mean lensing efficiency $\langle \beta \rangle$. We estimate the impact of cosmic variance using the \pz \ catalogues of \citet{coupon09} for the four CFHTLS DEEP fields. Again we analysed these fields with our own weak-lensing pipeline and matched these catalogues to introduce the lensing selections. These \pz's are based on five optical bands and hence are not as reliable as the COSMOS2015 catalogue. However, because the four fields were analysed consistently they may serve as an estimate of the variation in redshift distributions due to cosmic variance. For each cluster we compute the weighted average $\langle \beta \rangle$ for the 4 fields and use the standard deviation between them as the error due to cosmic variance. 

In addition to cosmic variance, there are Poisson errors in $\langle \beta \rangle$ due to finite statistics. The Poisson errors are estimated by comparing the lensing efficiency in the CFHTLS D2 field with the lensing efficiency in the remainder of the COSMOS field, where we assume that both regions of COSMOS have the same underlying distribution of galaxies. We compare the lensing efficiency for galaxies in the appropriate magnitude range for each cluster for both regions and use the difference as a measure of the Poisson error. As we do not have lensing measurements for the full COSMOS2015 catalogue we only impose the magnitude limits.

The previously mentioned \pz \ catalogues were all constructed using the LePHARE code \citep{ilbert06}. A final source of error we investigate is how different \pz \ algorithms change the mean lensing efficiency. For this we used the DR3 UltraVISTA catalogue (Muzzin et al., in prep) of 1.7 square degrees of the COSMOS field, constructed from the UltraVISTA survey, where sources were selected in K-band (see \citealt{hill17} for a description of the data). In the survey area there are stripes with extra deep observations covering 0.75 square degrees. The new DR3 catalogue is made using the same methods described in \citet{muzzin13} and \pz's are estimated with the EAZY code \citep{brammer08}. We redid our analysis with the DR3 catalogue and took the difference between $\langle \beta \rangle$ and our $\langle \beta \rangle$ from COSMOS2015 as the estimate for systematic uncertainties due to the algorithms.

We estimate our final uncertainty $\delta \beta$ by summing all three error sources quadratically, assuming they are independent. 
Cosmic variance is the dominant source of uncertainty, slightly higher than the redshift estimation and the Poisson error is negligibly small. The $\delta \beta$ estimates are listed in Table \ref{tbl:redshifts}. The uncertainty $\delta \beta$ is on average $\sim$2\%, but increases to 9\% for the highest redshift cluster, because the \pz's are more uncertain for the higher redshift objects in the COSMOS catalogue. Also, CCCP clusters have larger $\delta \beta$ values than \mn \ clusters due to the fainter source sample for CCCP.

\section{Contamination of the source population by cluster members}
\label{sec:contamination}

The galaxy catalogue from the lensing analysis contains both field galaxies and cluster members. Cluster members are not sheared by the gravitational potential of the cluster and keeping them in the sample will alter the shear signal. If cluster galaxies are not intrinsically aligned (indeed \citealt{sifon15} found no alignment), their presence dilutes the shear signal, biasing the shear estimate low, where the size of the bias depends on the relative overdensity of cluster members compared to background galaxies. 
Galaxies in front of the cluster also dilute the shear signal, but these are taken into account by the average critical surface density.

With reliable colours for individual galaxies, cluster members can be identified and removed from the sample \citep[e.g][]{medezinski18, varga19}. However, we lack the required multi-band observations. Instead, as was done by \citetalias{hoekstra15}, we apply a `boost~correction' to statistically correct for cluster member contamination. This approach offsets the dilution of the shear by boosting the shear signal based on the fraction of cluster members to background galaxies. The application of the boost correction relies on the assumption that only cluster members affect the galaxy counts. We investigate the effects that violate this assumption in the next sections and take them into account to obtain a reliable estimate of the density of cluster members relative to the  density of background galaxies, from which we compute the boost correction. 

As noted in Section~\ref{sec:data}, close proximity to bright objects can affect the measured shapes of galaxies, changing the measured shear signal. We incorporate this effect by quantifying the boost correction in terms of the sum of the lensing weights per square arcminute, which we call the weight density $\xi$. 
Here we only compute the boost corrections for \mn \ clusters and for CCCP cluster we use the corrections calculated in \citetalias{hoekstra15}.

\subsection{Magnification} 
\label{sec:magnification}
Gravitational lensing near the cluster core magnifies the background sky. This phenomenon increases the observed flux of background galaxies, but it also reduces the actual area behind the cluster that is observed. These two features counteract each other in their effect on the observed number density of sources. The net effect depends on the number of galaxies scattered into the magnitude range that we designate for our lensing study. The observed number of galaxies increases with the magnification $\mu$ as $\mu^{2.5 \alpha -1}$ \citep{mellier99}. Hence, for a slope of the magnitude distribution $\alpha = \mathrm{d log} N_\mathrm{source} / \mathrm{d} m_\mathrm{shape} = 0.40$ the net effect is negligible. For the MegaCam $r$-band data \citetalias{hoekstra15} computed that the slope is close to $0.40$ and so we can safely ignore the effect of magnification on the source population, especially for the data beyond 0.5 Mpc from the cluster centre.

\begin{figure}
 \centering
 \includegraphics[width=8.5cm,height=8.5cm,keepaspectratio=true]{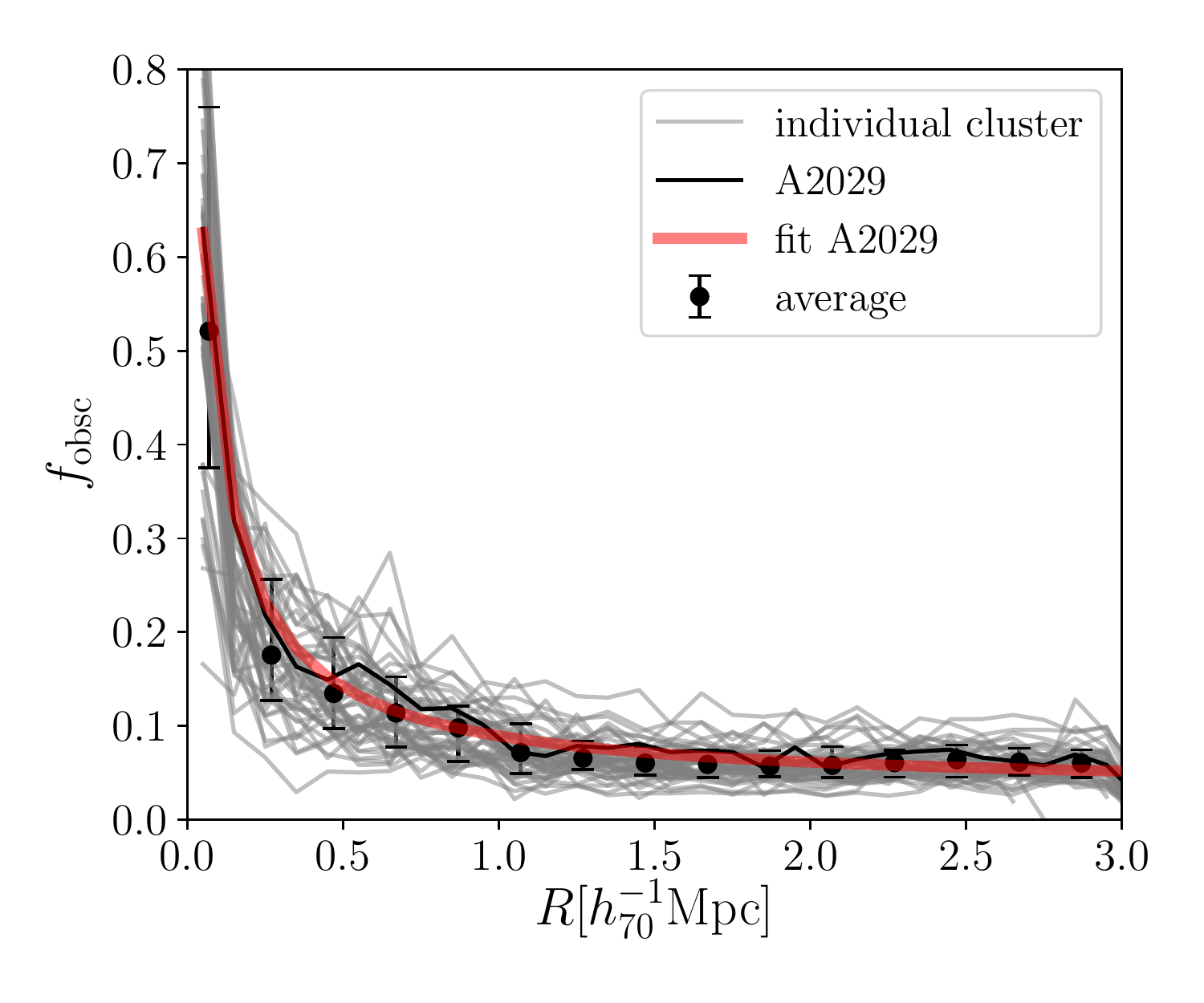}
 \caption{Obscuration of source galaxies by cluster members in realistic image simulations of \mn \ clusters as a function of radial distance to the cluster centre. Gray lines show the obscuration profile for individual clusters and the black points show the average for all clusters. The red line is an example of our fitting function to the obscuration profile of cluster A2029, which is shown as the black line. The region of interest for our lensing analysis is beyond 0.5 Mpc, where obscuration is on average only a few percent. 
 }
 \label{fig:obscuration}
\end{figure}

\subsection{Obscuration}
\label{sec:obscuration}
Cluster members are large foreground objects and obscure part of the background sky, thereby reducing the number density of observed background galaxies. This reduction affects our boost correction scheme. This phenomenon is especially important for \mn \ as the low redshift cluster members are large on the sky. To address this issue we use the results of \citet{sifon18}, who used image simulations of the \mn \ clusters to compute the effect of obscuration. Their cluster image simulations were designed to mimic the observations as closely as possible to accurately predict the effect of obscuration. For each simulated cluster image the seeing and noise level were set to the values measured in the data. Background galaxies were created with the image simulations pipeline of \citetalias{hoekstra15}, which is based on the \textsc{GalSim} software \citep{galsim}, and cluster galaxies were added to the images. \cite{sifon15} identified cluster members through spectroscopy or as part of the red sequence. Where available, the \textsc{GALFIT} \citep{galfit} measurements of \cite{sifon15} were used to create surface brightness profiles for galaxies. The distributions of measured \textsc{GALFIT} properties were then modeled with parametric curves. Some cluster members did not have (reliable) \textsc{GALFIT} measurements, and instead their properties were randomly sampled from these curves to create a surface brightness profile for the simulated images.
We ran the analysis pipeline on both the background image and the cluster image producing two lensing catalogues. By matching these catalogues, all background galaxies can be selected and the effect of cluster members on the weight density of the background population can be determined. We define obscuration as
\begin{equation}
 f_\mathrm{obsc} = 1 - \frac{\xi^\mathrm{cl}}{\xi^\mathrm{bg}} ,
 \label{eq:obscuration}
\end{equation} 
where $\xi^\mathrm{cl}$ and $\xi^\mathrm{bg}$ are the weight densities of all observed background galaxies in the cluster simulation and in the background simulation, respectively.

In Figure~\ref{fig:obscuration} we show the resulting obscuration in bins of projected cluster centric distance  $R$ for individual clusters in gray, and in black the average for all clusters. The effect of obscuration is greatest close to the cluster centre, which is expected because of the presence of the low redshift BCGs. At radii larger than 1 Mpc the obscuration flattens out but does not reach zero, even though we do not expect cluster members to obscure $\sim$5\% of all background galaxies in these outer regions. Instead, this plateau is caused by field galaxies entering the cluster member sample, as \citet{sifon15} showed that their sample of red sequence selected cluster members is contaminated at large radii. The simulated sample of cluster members lacks faint blue galaxies, but we expect that their obscuration is minimal over the range of interest: $0.5<R<2.0$ Mpc. Their addition to the obscuration would introduce a negligible contribution to the boost correction and we ignore them in our analysis.

We determine an obscuration correction for the background weight density in the \mn \ data by fitting a smooth function to the individual cluster obscuration profiles shown in gray in Figure~\ref{fig:obscuration}. We find that the expression
\begin{equation}
 f_\mathrm{obsc}(R) = n_\mathrm{\Delta} + n_\mathrm{0} \left( \frac{1}{R+R_\mathrm{c}} -  \frac{1}{R_\mathrm{max}+R_\mathrm{c}}  \right), 
 \label{eq:obsc_fit}
\end{equation}
worked well to describe the obscuration for $R<R_\mathrm{max}=3$ Mpc. The obscuration is set to $n_\Delta$ beyond $R_\mathrm{max}$. On average, $R_\mathrm{c} \approx 0.04$ Mpc and $n_0 \approx 0.04$ produce the best fits to the obscuration profiles. The parameter $n_\Delta$ was fit to capture the plateau at large radii. When creating the obscuration profile to be applied to the data, $n_\Delta$ is set to zero to renormalise the data such that $f_\mathrm{obsc}$ is consistent with zero beyond 1.5 Mpc. The best fits to the obscuration profiles to individual clusters were then used to correct the background galaxy counts in the \mn \ data.

\subsection{Excess galaxy weight density}
Now that we have a correction for the decreased weight density due to obscuration, we can determine the excess weight density of all sources in the \mn \ lensing catalogues relative to the weight density of background objects as a function of cluster-centric distance. This then provides the boost correction for the shear signal to correct for contamination of the source sample by cluster members.

The first step to compute the excess weight density is to determine the weight density of background objects. \citetalias{hoekstra15} used a halo model prediction to check that at 4 Mpc the structure associated to the cluster is a negligible contribution to the number density of field galaxies and used the area outside that 4 Mpc to estimate the field galaxy density. The low redshift of the \mn \ sample means that the field of view does not encompass 4 Mpc for all clusters. Only the highest redshift clusters have sufficient area outside 3 Mpc for statistically meaningful estimates. To compensate for this lack of data, we use ancillary publicly available observations of blank fields to obtain an estimate of the weight density of field galaxies $\xi^\mathrm{field}$ (as was also suggested by \citealt{schrabback18}). We selected 41 fields of deep CFHT data that do not contain clusters and have deeper imaging and have seeing values smaller than our observations. We analysed $\sim$33 square degrees of those fields with our lensing pipeline and we derive a parametric model for the field galaxy weight density $\xi^\mathrm{field}$ in Appendix~\ref{app:blanks}. The value of $\xi^\mathrm{field}$ is a function of the Galactic extinction, depth of the observations, and the seeing, and it predicts the mean density with an uncertainty of 1\%.
We use this model to predict the weight density of field galaxies for each cluster based on the seeing, depth and the Galactic extinction in the observations (listed in Table~\ref{tbl:redshifts}).

In the top panel of Figure~\ref{fig:excess_counts} we show the excess weight density $\xi/ \xi^\mathrm{field}$ (the obscuration corrected weight density normalised to the weight density of field galaxies), as a function of the distance to the BCG. Points with errorbars show the average excess weight density for all clusters and blue (red) shaded regions show the average excess weight density for clusters at $z<0.1$ ($z\geq0.1$). The contamination by cluster members is benign for the \mn \ clusters; the excess weight density is higher than 20\% only within the inner 500 kpc. For the lensing analysis we only use sources beyond 500 kpc (and sources beyond 2 Mpc are excluded due to mass modelling issues, see Section~\ref{sec:masses}), so the effect of contamination is small.

\subsection{Boost correction}
\label{sec:boostcorr}

\begin{figure}
 \centering
 \includegraphics[width=8.5cm,height=8.5cm,keepaspectratio=true]{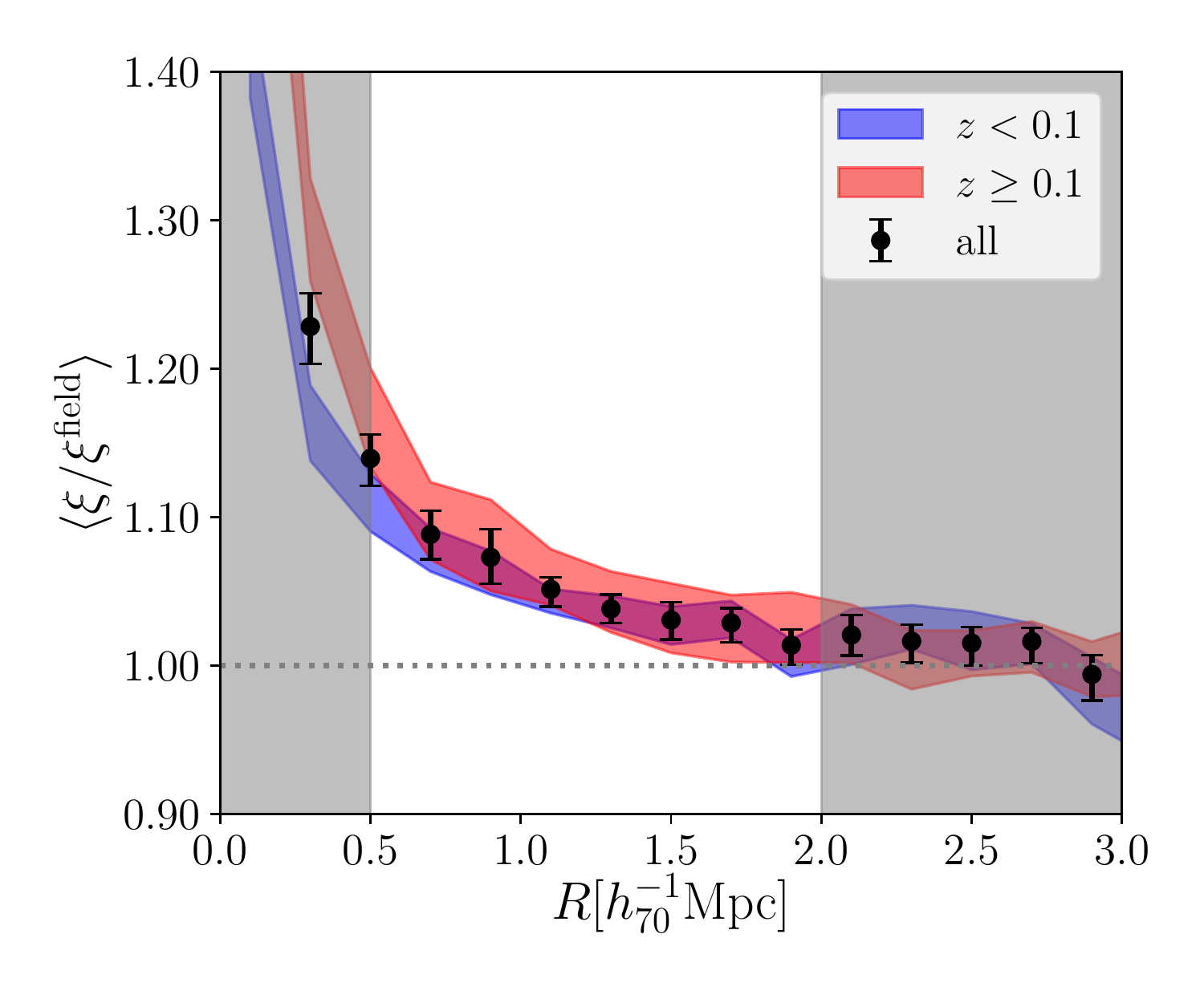}
  \includegraphics[trim={14 20 0 20},clip,width=8.5cm,height=8.5cm, keepaspectratio=true]{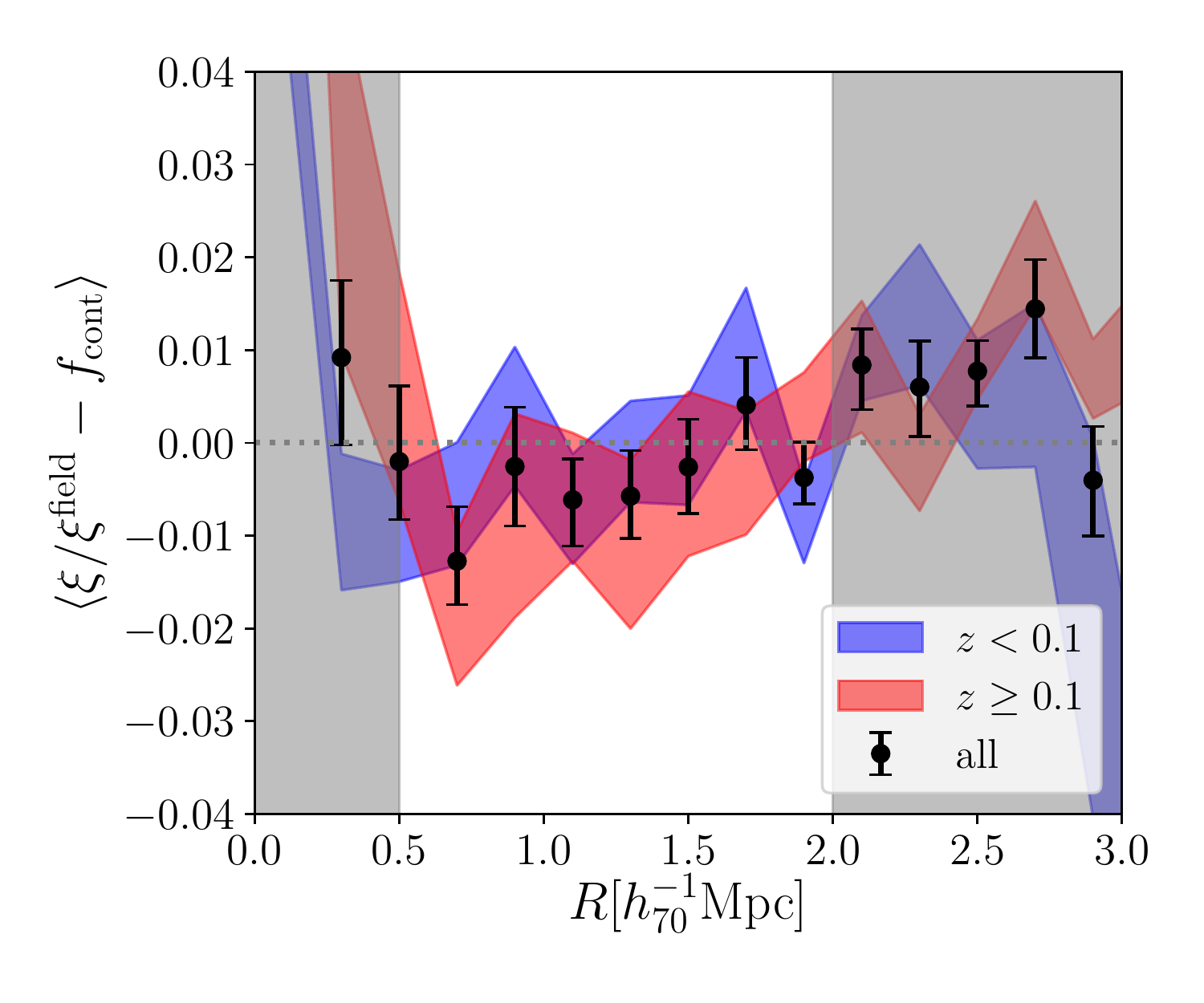}
 \caption{\textit{Top}: Excess weight densities of all sources in the magnitude range $20 \leq m_r \leq 24.5$ in the lensing catalogues as a function of radial distance to cluster centre. The excess weights are determined from the ratio of the weight density $\xi$ corrected for obscuration and the average weight density for field galaxies. Black points with errorbars show the average excess weight density for the full \mn \ sample, the blue (red) shaded area for all $z<0.1$ ($z\geq0.1$) clusters. The width of the coloured regions show the uncertainty on the mean excess weight density. The dotted line represents no contamination. The region shown in white between 0.5 Mpc and 2 Mpc is used for the lensing analysis in Section \ref{sec:nfw}, in which the contamination is on average $\sim$5\%. 
\textit{Bottom}: Same as top panel, but showing the average weight density of galaxies after the best fit model for contamination for each individual cluster has been subtracted. 
 }
 \label{fig:excess_counts}
\end{figure}

The excess weight density per cluster is a noisy measurement and using it directly to boost the shear signal can produce a spurious signal. Instead, we assume that the density of cluster members is a smooth function of the cluster-centric radius. This assumption will not be valid if the cluster has local substructure, but any additional uncertainty this introduces will average out for the full ensemble of clusters. Like \citetalias{hoekstra15} we use Equation~\ref{eq:obsc_fit}, where the amplitude of the contamination $n_0$ and the cluster core radius $R_\mathrm{c}$ are fitted for each cluster individually. The maximum radius $R_\mathrm{max} = 3$ Mpc is the limit beyond which the function is set to $n_\Delta$. In Figure \ref{fig:excess_counts} the excess weight density already vanishes beyond 2 Mpc, so setting $R_\mathrm{max} = 3$ Mpc is reasonable for \mn. All CCCP clusters were small enough in angular coordinates so that \citetalias{hoekstra15} could set $n_\Delta=1$. However, our prediction for field galaxy weight density has an intrinsic scatter and so we do not expect the excess weight density for individual clusters to converge to 1 at large radii. Therefore we add $n_\Delta$ as a free parameter in our analysis. We find that the relative spread in $n_\Delta$ is 7.2\%, which is in agreement with the 6.4\% scatter expected from the blank fields.

The ensemble averaged residual, after subtracting the best fit profile for each cluster from its excess weight density, is shown in the bottom panel of Figure~\ref{fig:excess_counts}. Again, we separate the sample in low redshift ($z<0.1$, blue) and high redshift ($z\geq0.1$, red) clusters and the full sample is denoted by the black points. 
For most radii the average residual is consistent with zero within the errors, regardless of the mean redshift of the sample. This shows that Equation~\ref{eq:obsc_fit} is a decent description of the density of cluster members. At $R \approx 3$ Mpc the observed area for $z<0.1$ clusters is decreasing which greatly increases the errorbars and the crowded cluster centre is not accurately described by the fitting function. However, for the lensing analysis in Section~\ref{sec:nfw} we restrict ourselves to 0.5 - 2 Mpc for which the residual is consistent with zero with an uncertainty of $\sim$1.5\%. The best fit profiles will serve as a boost correction for the shear signal of clusters to statistically correct for contamination of the source population by unlensed cluster members.

\section{\mn \ cluster masses}
\label{sec:masses}

In the previous sections we have computed the corrections owing to the lack of individual redshift estimates for the source galaxies and the presence of cluster members in the source sample. We now apply these corrections to the measured tangential shear and use the resulting shear as a function of cluster-centric distance to estimate the weak-lensing masses using two different methods. Only data beyond 0.5 Mpc are used in the mass calculations, because the shear calibration was not tested for large shear values (Section~\ref{sec:data}), and the (residual) contamination is small far from the cluster centre (Section~\ref{sec:contamination}). In addition, this radial cut reduces the impact of miscentring (see Section~\ref{sec:mass_error}).

The mass modelling pipelines described in the next two sections may not perfectly recover the cluster mass. We check the accuracy of the pipelines with the state-of-the-art HYDRANGEA cluster simulations \citep{bahe17, barnes17}, finding that our masses are underestimated by only 3-5\%. The details of this analysis can be found in Appendix~\ref{app:massbias}. To account for scatter due to uncorrelated structures along the line of sight \citep{hoekstra01}, we use predictions from \citet{hoekstra11} to incorporate the effect into the errorbars on our weak-lensing masses.

\subsection{Navarro-Frenk-White profile}
\label{sec:nfw}

An often used profile to describe dark matter haloes is the Navarro-Frenk-White (NFW) profile, which is known to be a good fit to observational data \citep[e.g.][]{okabe13, umetsu14, viola15}. In numerical simulations \citet{navarro97} found a universal profile for the density of dark matter haloes
\begin{equation}
 \frac{\rho (r)}{\rho_0} = \frac{\delta_c(\Delta)}{(r/r_\mathrm{s}) (1+r/r_\mathrm{s})^2} ,
 \label{eq:nfw}
\end{equation}
where the radial shape of the profile is defined by the scale radius  $r_\mathrm{s}$\footnote{Here the radius $r$ and the scale radius $r_\mathrm{s}$ are three dimensional quantities in contrast to the capitalised two dimensional cluster-centric radius $R$.}. The amplitude of the profile is set by the characteristic overdensity 
\begin{equation}
\delta_c(\Delta) = \dfrac{\Delta}{3} \dfrac{c_\Delta^3}{\mathrm{ln}(1+c_\Delta) +c_\Delta/(1+c_\Delta)} , 
 \label{eq:deltaconc}
\end{equation}
which depends on the concentration $c_\Delta$. For a fixed number $\Delta$, the concentration $c_\Delta$ is the ratio of the radius $r_\Delta$ enclosing a sphere of density $\Delta \hspace{1pt} \rho_0$ and the scale radius: $c_\Delta = r_\Delta/r_\mathrm{s}$. The mass within this region can be obtained from:
\begin{equation}
 M_\Delta = M(\Delta, r_\Delta) = \Delta \hspace{1pt} \rho_0  \frac{4 \pi}{3} r_\Delta^3 .
 \label{eq:nfwmass}
\end{equation}
The density $\rho_0$ is usually set to the critical density of the Universe $\rho_\mathrm{crit} = 3 H(z)^2/8 \pi G$.

We follow the definitions in \cite{wright00} to fit a projected NFW profile to our lensing signal. We combine their expressions for $\gamma$ and $\kappa$ to create an NFW profile for the tangential reduced shear $g$, again with the additional terms given in Equation~\ref{eq:photzbias}. The free parameters in the NFW model are correlated and the concentration depends on redshift. In practice, the concentration is constrained using numerical dark matter simulations. We follow \citetalias{hoekstra15} and use the mass concentration relation found by \citet{duttonmaccio14}, which is in agreement with later work \citep{diemerkravtsov15}. With the addition of the mass-concentration relation, our fitting function only has the mass $M_\Delta$ as a free parameter. 
The scales at which we fit our NFW model are restricted to 0.5 - 2 $h_{70}^{-1}$ Mpc scales, because at large radii the two-halo term begins to dominate the signal. For the nearest clusters the field of view is not large enough to reach 2 Mpc and instead we take an outer radius of $1500''$.
We compute the mass at overdensities of 200 and 500 times $\rho_\mathrm{crit}$, $M_{200}$ and $M_{500}$, respectively. In Appendix \ref{app:massbias} we compute the ratio of our mass estimates from NFW fitting and the true mass using simulations and find it to be 0.93 and 0.97 at $r_{200}$ and $r_{500}$, respectively. The masses computed for the \mn \ and CCCP cluster masses with our pipeline are divided by these values and the corrected masses are listed in Table~\ref{tbl:masses}.

\begin{figure}
 \centering
 \includegraphics[width=8.5cm,height=8.5cm,keepaspectratio=true]{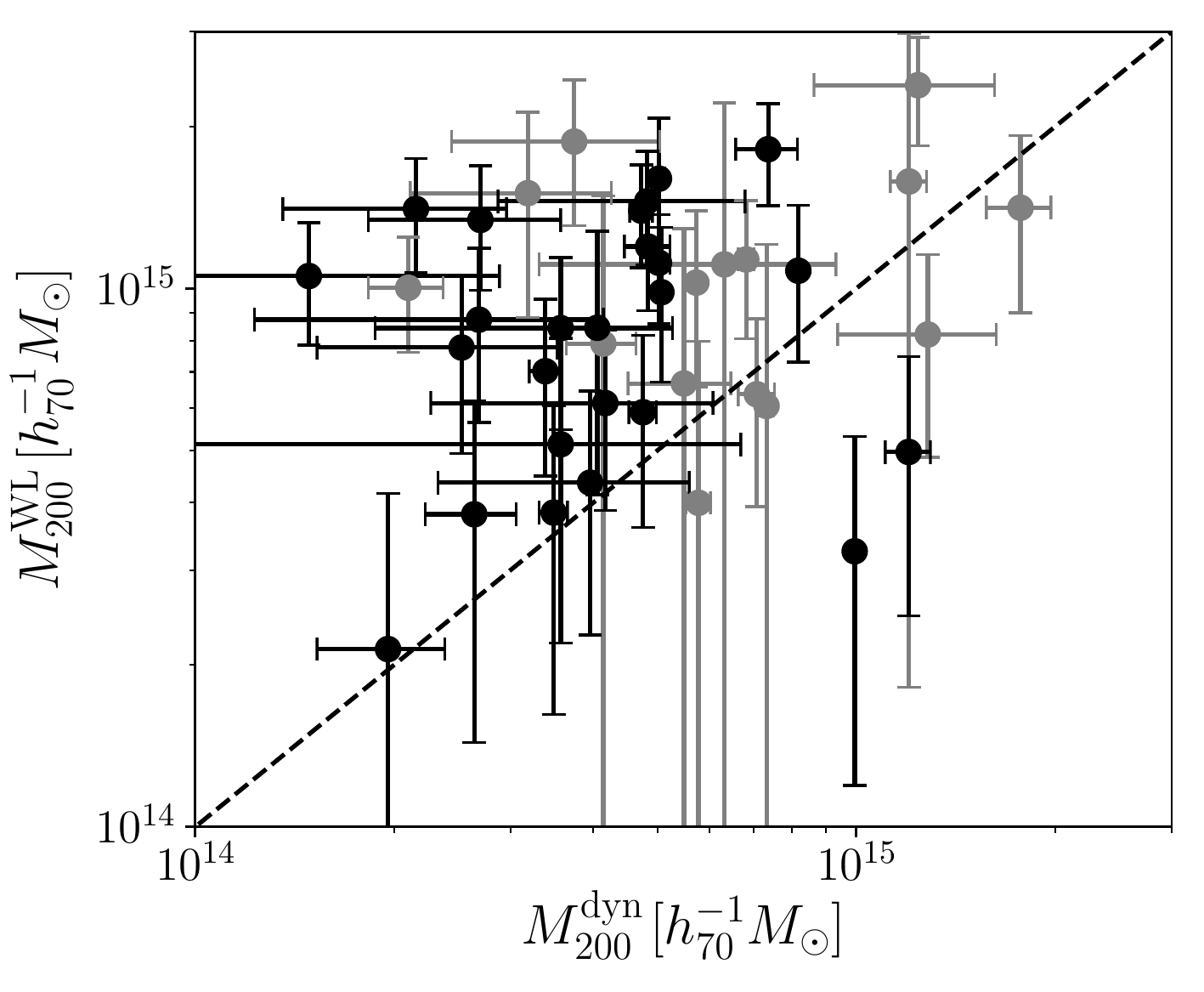}
 \caption{Comparison of the weak-lensing masses $M^\mathrm{WL}_{200}$ and the dynamical caustic masses $M^\mathrm{dyn}_{200}$ from \citet{rines16}.  Black points show our results and gray points show the results for CCCP clusters from \citetalias{hoekstra15}. The dashed line shows unity slope. }
 \label{fig:rinescomp}
\end{figure}

It is instructive to compare our best fit NFW masses to other available mass estimates. We discuss one comparison here and discuss other weak-lensing measurements in Section~\ref{sec:hsebias}. \citet{rines16} have used spectroscopic redshifts to identify caustics in the phase-space distribution of member galaxies, which can be related to the escape velocity in the cluster potential. They provide $M_{200}$ dynamical masses for 25 \mn \ clusters and 15 CCCP clusters and the comparison to our lensing estimates is shown in Figure~\ref{fig:rinescomp}.
It is clear that the weak-lensing masses are generally higher than the dynamical masses. This discrepancy is consistent for both the \mn \ and the CCCP sample. \citetalias{hoekstra15} discussed that the discrepancy could be reduced (but not removed) for CCCP by excluding outliers that were commented upon by \citet{rines13}. The bulk of the \mn \ clusters have consistently higher weak-lensing mass compared to the dynamical mass, making it difficult to explain the difference based on individual clusters. We also find no correlation between the state of relaxedness of the clusters (see Section~\ref{sec:hsebias}) and the difference in caustic and weak-lensing mass. The discrepancy is much larger than the several percent level systematic errors we have computed for the lensing masses. We could not find a satisfactory explanation for the discrepancy of the mass estimates, but we note that dynamical masses can suffer from large biases and scatter \citep{old15, old18,armitage18}.

\subsection{Aperture masses}
\label{sec:apmass}

An alternative to fitting density profiles to the data is to directly measure the mean convergence in an aperture of radius $R_1$ relative to the density in an annulus at $R_2$ and $R_\mathrm{max}$ using the expression
\begin{align}
  & \overline{\kappa}(R\leq R_1) - \overline{\kappa}(R_2 < R \leq R_\mathrm{max}) = \nonumber \\ 
  & 2 \int_{R_1}^{R_2}  \langle \gamma_\mathrm{t} \rangle d\mathrm{ln}R + 2 \frac{R^2_\mathrm{max}}{R^2_\mathrm{max}-R^2_2} \int_{R_2}^{R_\mathrm{max}}  \langle \gamma_\mathrm{t} \rangle d\mathrm{ln}R &
  \label{eq:apmass}
\end{align}
\citep{clowe98}. This relation gives a direct measurement of the mean surface mass density, but requires knowledge of the mean convergence in the annulus and the tangential shear profile, both of which are unknown. Fortunately, these can be estimated using the convergence profile of the best fit NFW profile. Far from the cluster centre the convergence will be small, so the difference between shear $\gamma$ and reduced shear $g$ should be negligible, and if $R_2$ is chosen far from the aperture radius $R_1$, the contribution of the annulus should be modest and any bias from the assumption of the NFW profile small.

In practice, the low redshift of the \mn \ clusters limits the physical values of $R_2$ and $R_\mathrm{max}$ that will fit inside the MegaCam field of view. We choose an outer radius $R_\mathrm{max}=1500''$ based on the degradation of quality of the observations outside that radius. This corresponds to 1.3 Mpc, 2.7 Mpc, and 3.9 Mpc at $z=$ 0.05, 0.10, 0.15, respectively. $R_2$ has to be chosen far enough away from $R_1$ to reduce the impact of the assumption of an NFW profile for $\overline{\kappa}(R_2 < R \leq R_\mathrm{max}) $, but it must also not be to close to $R_\mathrm{max}$ to avoid large uncertainties in the integral from $R_2$ to $R_\mathrm{max}$ in Eq. \ref{eq:apmass}. We set \mbox{$R_2 = 900''+400''(0.15-z_\mathrm{cl})$} so that the lowest redshift clusters at $z_\mathrm{cl} \approx 0.05$ have the thinnest annuli, allowing for measurements around $R_1=$1 Mpc. For CCCP we do not alter the $R_2$ and $R_\mathrm{max}$ values from \citetalias{hoekstra15}: $600''$ to $800''$ for clusters observed with CFH12k and $900''$ to $1500''$ for clusters observed with MegaCam. The $R_2$ and $R_\mathrm{max}$ values used for each cluster in physical units are listed in Table~\ref{tbl:redshifts}.

\begin{figure}
 \centering
 \includegraphics[width=8.5cm,height=8.5cm,keepaspectratio=true]{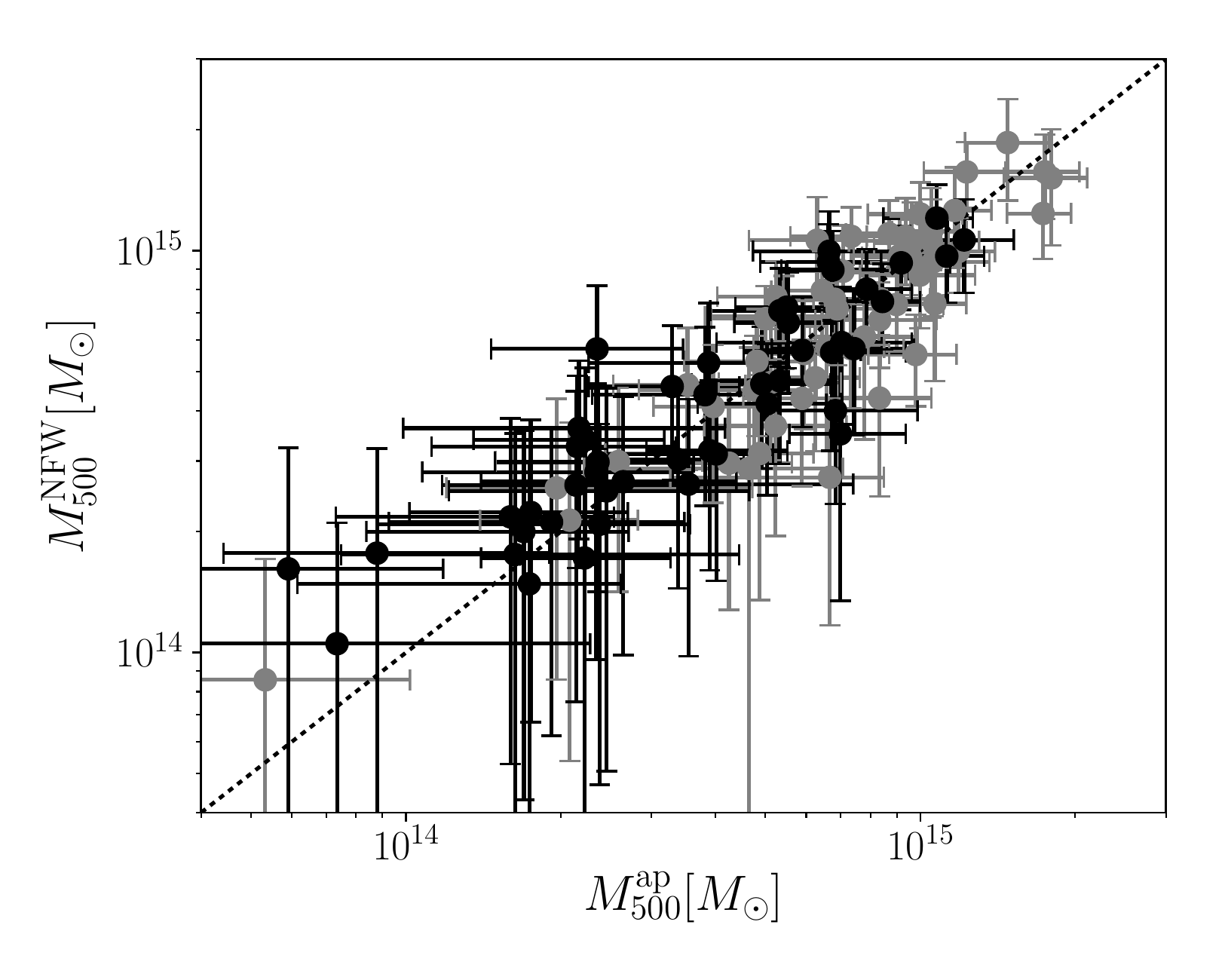}
 \caption{ Comparison of the $M_{500}$ mass measured with the deprojected aperture method and NFW fitting for \mn \ clusters in black and CCCP clusters in gray. The line shows equality.  }
 \label{fig:mapcomp}
\end{figure}

A drawback of Eq. \ref{eq:apmass} is that it only provides a measure for the projected mass, whereas most other mass proxies are calculated inside a sphere. To deproject the aperture mass estimates we assume that the matter along the line of sight is distributed as an NFW profile. 
In practice we first find the NFW mass $M_\Delta$ (again with the \citealt{duttonmaccio14} mass-concentration relation) that reproduces the mean convergence $\overline{\kappa}(R \leq R_1)$ measured from the data with Eq \ref{eq:apmass}. Then for that NFW profile we calculate the spherically enclosed mass at $R_1$ as the deprojected aperture mass. We repeat this for the range of $R_1$ for which we measured the 2D mean convergence. We interpolate between measurements to find $r_{500}$. The radius $r_{200}$ is larger than the available field of view for many clusters and even $r_{500}$ is barely in view for the nearest clusters. 
 In Appendix \ref{app:massbias} we determine the bias in aperture masses using the HYDRANGEA cluster simulations and find that masses are underestimated by $\sim$2\% for most of the sample. The $M_{500}$ and $r_{500}$ estimates are corrected accordingly and listed in Table \ref{tbl:masses}.
The deprojected aperture masses are in reasonable agreement with the NFW masses (see Figure \ref{fig:mapcomp}) for our cluster sample. A simple linear fit with bootstrap errors shows that $M^\mathrm{NFW}_{500} / M^\mathrm{ap}_{500}$=0.98$\pm$0.03. This is in good agreement with our results from the HYDRANGEA simulations.

\subsection{Systematic error budget}
\label{sec:mass_error}

A large part of this work has been devoted to corrections for systematic effects. Here we review their impact on our mass estimates.

\begin{itemize}
\item In our analysis we have assumed that the centre of the cluster is given by the location of the BCG. If the BCG is not at the bottom of the gravitational potential the mass estimates will be biased. However, \citet{hoekstra11b} show that for our conservative choice of 0.5 Mpc as the lower limit of the fit range the bias is only $\sim$5\% if the BCG is 100 kpc from the true cluster centre. If the distance between the BCG location and the peak in the X-ray surface brightness is small, they are a good indicator of the centre of the gravitational potential of the cluster \citep{george12}. \citet{mahdavi13} and \citet{bildfell13} found that most of the CCCP clusters have a BCG offset smaller than 100 kpc. \citet{lopes18} also find the distance between the X-ray peak and the BCG for a dozen of the \mn \ clusters to be $\lesssim$ 100 kpc, with only 4 BCGs further than 10 kpc off from the X-ray peak.  We thus expect a miscentring bias to be negligibly small compared to our statistical errors for our mass estimates.

 \item The uncertainty in the shear estimates for our pipeline was tested by \citetalias{hoekstra15} and they found an accuracy of $\sim$1\%. They conservatively assign a 2\% uncertainty in their analysis and we do the same. 
 
 \item Thanks to the new high-fidelity COSMOS2015 \pz \ catalogue, the uncertainty in our source redshift distribution is on average 2\%. For \mn \ it is $\lesssim$1\% and for CCCP it is between 2\% and 9\%, increasing with cluster redshift, due to the increase in uncertainty for faint and distant galaxies in our utilized \pz \ catalogues.
 
 \item The boost corrections applied to our tangential shear profiles are accurate to $\sim$1.5\%.
 
 \item The uncertainty on the mean mass deduced from cluster simulations is $\sim$3\% for NFW masses and $\sim$2\% for aperture masses.
\end{itemize}

We treat these sources of errors as uncorrelated and we add them quadratically to find an average systematic error of 4.5\%. This value decreases for low redshift clusters and when using the aperture masses. The dominant error sources are the mass modeling, which can be improved with more simulations tailored to our selection of clusters, and the photometric redshift distribution.

\section{Comparison with SZ and X-ray}
\label{sec:hsebias}

The \textit{Planck} all-sky survey has produced a large catalogue of clusters detected through the SZ effect \citep{planckSZclusters16}. \citet{planckclustercosmo16} used 439 clusters to constrain cosmological parameters by measuring the cluster mass function. Cluster masses were computed using a scaling relation between the hydrostatic X-ray mass and the SZ observable $Y_\mathrm{SZ}$ (integrated Compton $y$-profile) based on a pressure profile, calibrated using measurements of 20 nearby relaxed clusters \citep{arnaud10}. X-ray mass estimates can be biased because the underlying assumption of hydrostatic equilibrium is violated in galaxy clusters by bulk gas motions and non-thermal pressure support \citep[e.g.][]{rasia12}, or due to uncertainties related to the calibration of X-ray temperature \citep{mahdavi13, schellenberger15}, and possibly by the assumption of a pressure profile. \citet{planckclustercosmo16} find that a bias $M_{Planck}/M_\mathrm{true} \equiv 1-b = 0.58 \pm 0.04$ is required to attain consistency between cosmological parameter constraints obtained with the cluster mass function and those obtained using primary CMB measurements \citep{planckcosmo16}. Such a low bias is not fully supported by independent weak-lensing mass measurements. \citet{vonderlinden14b} find $1-b = 0.69 \pm 0.07$, which is consistent with $0.58 \pm 0.04$, but \citetalias{hoekstra15} find a higher value $0.76 \pm 0.05 \mathrm{(stat)} \pm 0.06 \mathrm{(syst)}$. \citet{pennalima16} found $0.73 \pm 0.10$, and \citet{smith16} and \citet{gruen14} find that $1-b$ is consistent with one. Recent measurements from the Hyper Suprime-Cam Survey found $1-b=0.80\pm{0.14}$ \citep{medezinski18b} and $1-b=0.74^{+0.13}_{-0.12}$ \citep{miyatake19}. The recent re-analysis of \textit{Planck} CMB cluster lensing has found $1-b=0.71 \pm 0.10$ by \citet{zubeldia19}. However, \citet{battaglia16} showed that adding a correction for Eddington bias would move the results of WtG and CCCP more in line with the required value for consistency.

\begin{figure}
 \centering
 \includegraphics[width=8.5cm,height=8.5cm,keepaspectratio=true]{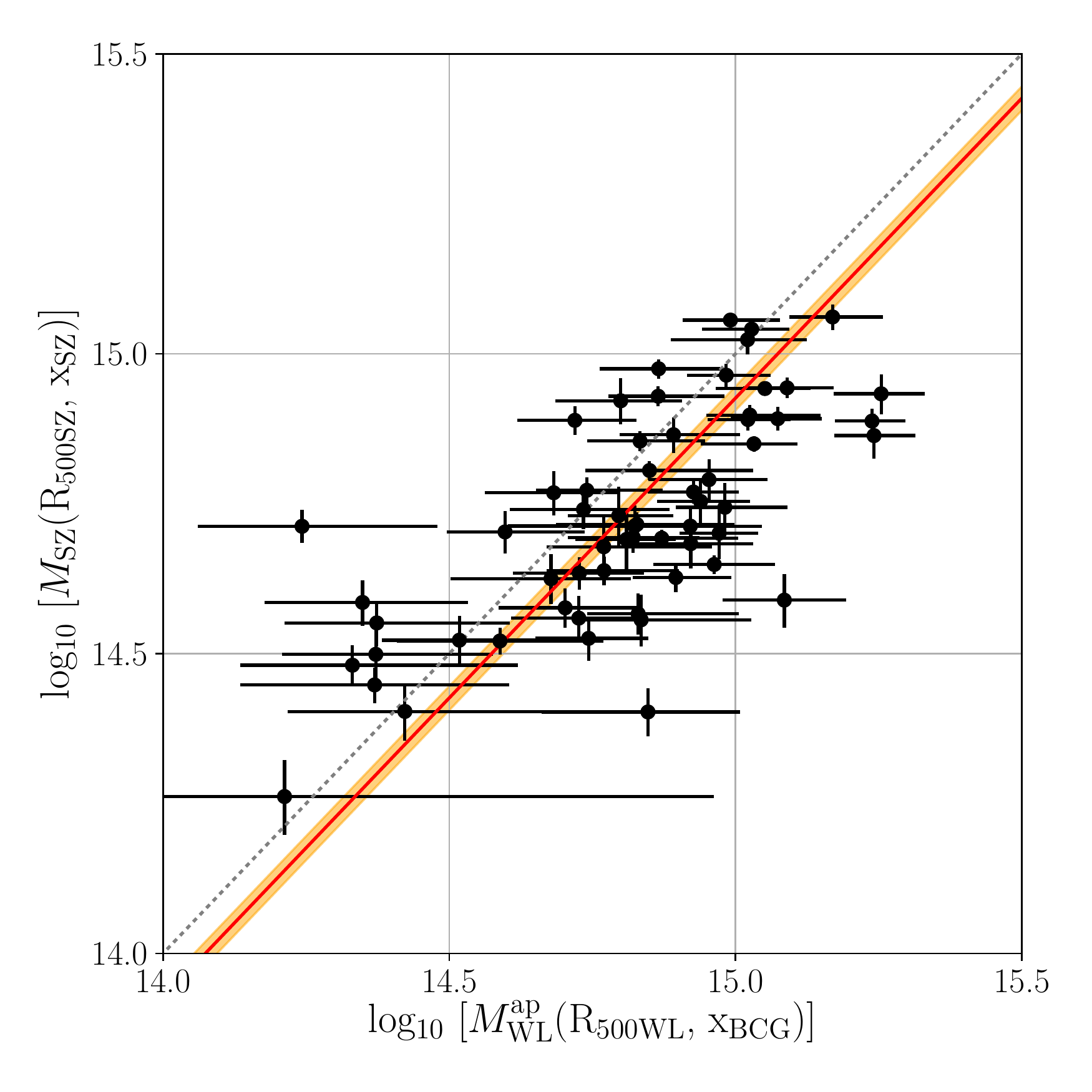}
 \caption{Comparison of the weak-lensing masses $M_{500}$ and the re-extracted SZ masses from \textit{Planck} for 61 clusters.  Black points show our results and the red line shows the best fit scaling relation using a constant hydrostatic mass bias, with $1 \sigma$ uncertainty shown as the orange band. The dotted line shows a one-to-one relation. 
 }
 \label{fig:planckcomp}
\end{figure}

In Figure~\ref{fig:planckcomp} we show our weak-lensing aperture mass measurements $M^\mathrm{ap}_\mathrm{WL} (R_\mathrm{500WL}, \mathrm{x_{BCG}})$ within the weak-lensing derived $R_\mathrm{500WL}$, centered on the BCG position $\mathrm{x_{BCG}}$, as our best estimate of the total cluster mass, and the SZ masses $M_\mathrm{SZ} (R_\mathrm{500SZ},  \mathrm{x_{SZ}})$ based solely on the \textit{Planck} measurements of $R_{500}$ and the cluster center. The \textit{Planck} SZ masses were extracted using the MMF3 pipeline \citep{planckSZclusters16}. 
Not shown in the figure are A115N, A115S, A223N and A223S from our sample, because A115 and A223 were measured as single clusters by \textit{Planck}. Like \citetalias{hoekstra15}, we also omit A2163 from the sample. We fit 61 clusters to constrain $1-b \equiv M_\mathrm{SZ} / M_\mathrm{WL}$. We use the {\tt LRGS} {\tt R}-package \citep{mantz16b} to perform the fit allowing for intrinsic scatter. The aperture mass measurements are taken as the weak-lensing masses, but similar results are obtained when using the NFW masses. 
The best fit value is $1-b = 0.84 \pm 0.04$ with an intrinsic scatter of $28 \pm 5$\% in the lensing mass at a given SZ mass. This relation is shown in Figure~\ref{fig:planckcomp} as the red line. This value of the mass bias is somewhat higher, but consistent with most weak-lensing results from the literature. The intrinsic scatter is expected to be dominated by the scatter in the weak-lensing mass due to the triaxial nature of dark matter haloes \citep[e.g.][]{herbonnet19} and our result is consistent with the scatter found in simulations \citep[e.g.][]{meneghetti10, henson17}. However, we note that both lensing and SZ measurements are sensitive to halo orientation and this correlation will lower the measured intrinsic scatter.

Selection effects can strongly affect the inferred scaling relation \citep[e.g.][]{mantz19}.
However, the selection function for our combined sample of \mn \ and CCCP is not trivial and we do not attempt to incorporate selection biases into our analysis. To investigate the effect of different selection criteria we introduce various selections and then remeasure the mass bias. Given our large sample these selections do not result in a significant loss of statistical power. The results of the different selections are summarized in Table~\ref{tbl:selbias}.

In our tests with cluster simulations (Appendix~\ref{app:massbias}) a massive system of two merging galaxy clusters in the plane of the sky has a total mass underestimated by a factor of $\sim$1.4. To identify such very-unrelaxed systems in our sample we used the symmetry, peakiness, and alignment X-ray measurements from \citet{mantz15b}, who used these to determine the relaxedness of their clusters. Clusters with high values for these parameters are deemed relaxed (see their Figure~8) and we call clusters with low values of symmetry and alignment \textit{very disturbed}; we found no added benefit from including peakiness measurements for our selection. Known major mergers in our full sample, such as A2163 and A520, fall within this very disturbed category, but our classification does not capture all known mergers (e.g. A754). We then removed these very disturbed clusters from the full sample and find $1-b=0.81\pm0.05$ and an intrinsic scatter of $29\pm6$\%. The value of $1-b$ is consistent with the results from the full sample and surprisingly we see no effect on the intrinsic scatter.

We imposed SNR cuts on the measured $Y^\mathrm{SZ}$ observable. We find that $1-b$ increases as we impose higher SNR cuts. Applying SNR cuts to the \textit{Planck} observable can lead to a change in $1-b$ of up to 1$\sigma$ (of the order of 0.04). 
We also mimicked an X-ray selected sample by imposing a limit on the X-ray flux $F_X$. For this we matched our results to the MCXC catalogue \citep{piffaretti11}, where 7 clusters were not matched. We find a change of 0.03 from $1-b=0.84\pm0.04$ when imposing the flux limit of the REFLEX survey \citep{bohringer04} of $3\cdot10^{-12} $ergs/s/cm$^2$ and a maximum shift of 0.07 when increasing this flux limit by a factor of 1.5, 2.0 or 3.0.

\begin{table}
\begin{tabular}{l c c l}
selection &$N_\mathrm{c}$ &$1-b$ & \multicolumn{1}{c}{IS} \\
\hline
all  & 61 & $0.84 \pm 0.04$ &  $28 \pm 5$\%  \\
no very disturbed clusters & 43 & $0.81 \pm 0.05$ &  $29 \pm 6$\%  \\
SNR $Y_\mathrm{SZ}>7.0 $  & 49 & $0.85 \pm 0.04$ &  $30 \pm 6$\%  \\
SNR $Y_\mathrm{SZ} >8.0 $  & 39 & $0.87 \pm 0.05$ &  $29 \pm 7$\%  \\
SNR $Y_\mathrm{SZ} >10.0 $  & 29 & $0.88 \pm 0.06$ &  $31 \pm 8$\%  \\
SNR $Y_\mathrm{SZ} >13.0 $  & 20 & $0.86 \pm 0.06$ &  $25 \pm 14$\%  \\
$F_X>3.0\cdot10^{-12}$ ergs/s/cm$^2$ & 54 & $0.87 \pm 0.06$ &  $30 \pm 7$\%  \\
$F_X>4.5\cdot10^{-12}$ ergs/s/cm$^2$ & 44 & $0.88 \pm 0.05$ &  $27 \pm 6$\%  \\
$F_X>6.0\cdot10^{-12}$ ergs/s/cm$^2$ & 36 & $0.87 \pm 0.06$ &  $30\pm 7$\%  \\
$F_X>9.0\cdot10^{-12}$ ergs/s/cm$^2$ & 24 & $0.91 \pm 0.07$ &  $25\pm 8$\%  \\
\\
selection & $N_\mathrm{c}$ &$f_\mathrm{gas}$ & \multicolumn{1}{c}{IS} \\
\hline
all & 41 & $0.130 \pm 0.006$ & $36 \pm 16$\% \\ 
no very disturbed clusters & 27 & $0.126 \pm 0.007$ & $19 \pm 16$\% \\
relaxed clusters & 8   & $0.139 \pm 0.014$ & $24 \pm 21$\% \\
\end{tabular}
 \caption{Results from the scaling relation analysis for different selections of our cluster sample described in the text and the number of clusters in that selection, $N_\mathrm{c}$. Comparison of weak-lensing masses and the SZ masses from \textit{Planck}, $1-b = M_{Planck}/M_\mathrm{WL}$, shown in the top section, and weak-lensing masses and the gas masses,, $f_\mathrm{gas}= M_{gas}/M_\mathrm{WL}$,  from \citep{mantz16} in the bottom section of the table, together with the intrinsic scatter in the weak lensing masses (IS). 
 }
 \label{tbl:selbias}
\end{table}

\begin{figure}
 \centering
 \includegraphics[width=8.5cm,height=8.5cm, keepaspectratio=true]{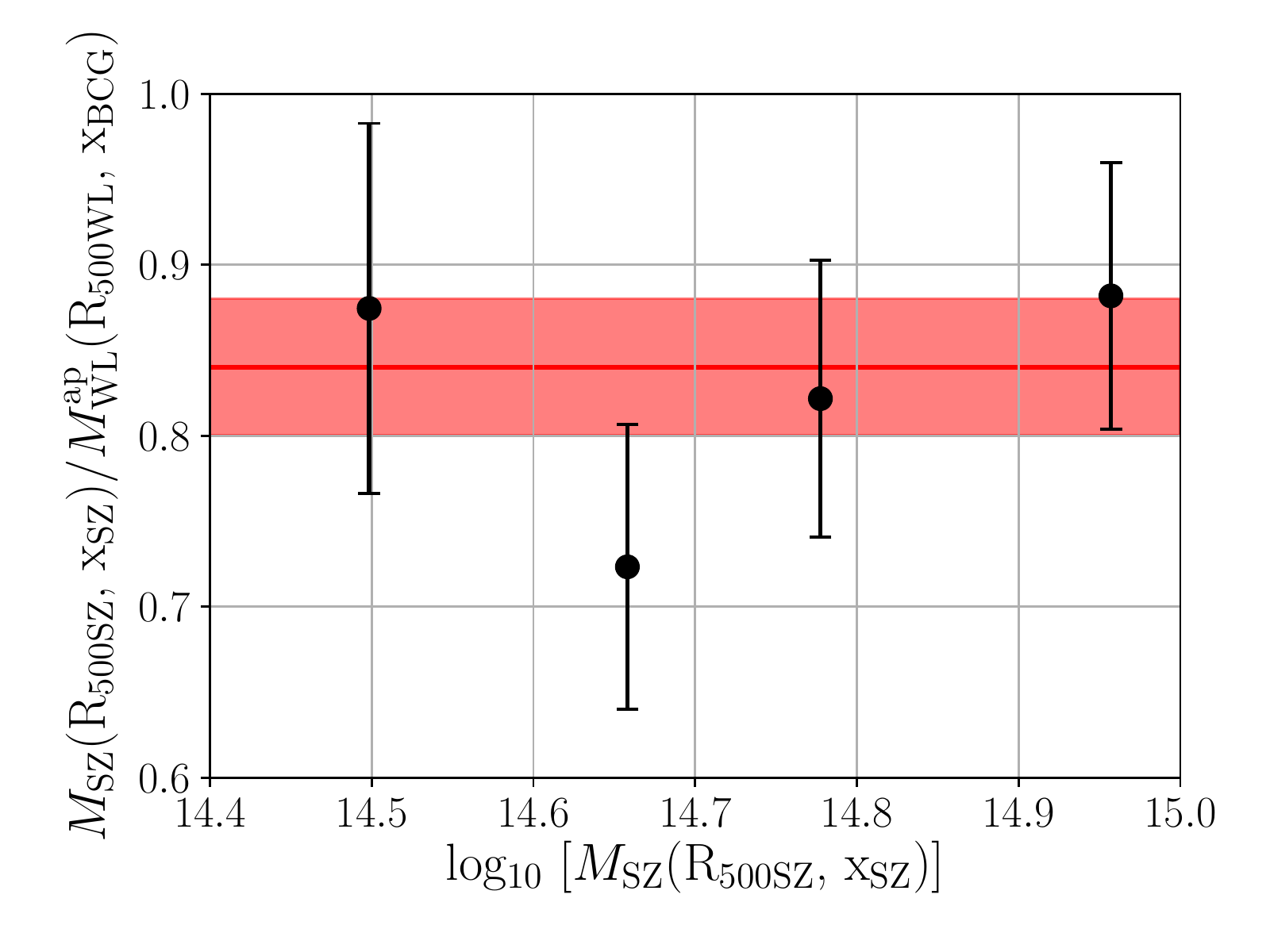}
 \includegraphics[width=8.5cm,height=8.5cm, keepaspectratio=true]{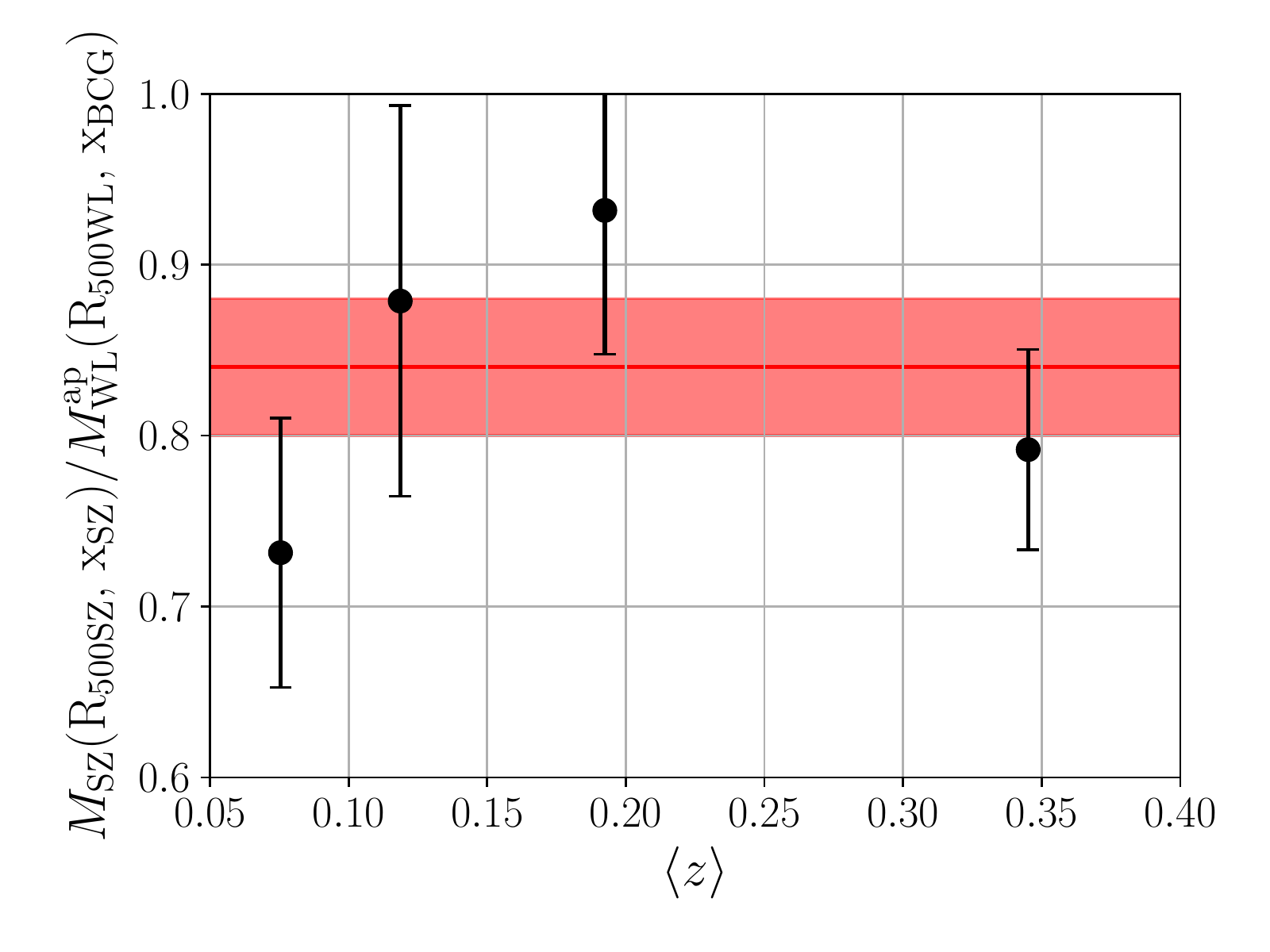}
 \caption{Best fit SZ mass bias in bins of SZ mass (\textit{top}) and bins of redshift (\textit{bottom}) for 61 clusters, both showing no significant trend within the uncertainties. Each bin contains roughly equal numbers of clusters. The red line and area show the best fit results and 1$\sigma$ uncertainty for the full sample.
 }
 \label{fig:biasinbins}
\end{figure}

Several observations \citep[\citealt{vonderlinden14b}; \citetalias{hoekstra15}; ][]{mantz16, eckert19} and simulations \citep{henson17} show that $1-b$ changes with cluster mass. To investigate any trend of $1-b$ with mass, we divide our cluster sample into four bins in SZ mass with roughly equal numbers of clusters and repeat our analysis for each bin. The resulting mass biases are shown in the top panel of Figure~\ref{fig:biasinbins}. We see no significant trend with mass for our clusters within the large uncertainties.
Alternatively, \citet{smith16} and \citet{gruen14} saw a redshift dependence of the mass bias in their cluster sample, as did \citet{salvati19} in their cosmological analysis of \textit{Planck} clusters, possibly arising due to systematic errors in weak-lensing measurements or departures from self-similar evolution. The selection of the clusters in our sample is loosely based on a flux limited survey, so we expect mass and redshift to be correlated. However, since we see no mass trend, we checked for a redshift trend. The result is shown in the bottom panel of Figure~\ref{fig:biasinbins} and there is no significant redshift dependence of the bias. 

These tests suggest that different selections can lead to a change of up to 0.07 in $1-b$ for our analysis. However, a proper determination of the scaling relations including selection bias requires a more careful analysis \citep{mantz16, bocquet19}.

We also determine the scaling relation between gas mass and weak-lensing mass at $r_{500}$ to obtain the gas fraction $f_\mathrm{gas}=M_\mathrm{gas}/M_\mathrm{WL}$, which is a cosmological probe \citep[e.g.][]{mantz14}. We use the gas masses presented in \citet{mantz16} for 42 clusters in our sample and measure the weak-lensing masses within the $r_{500}$ estimates from  \citet{mantz16} and around their assumed centres. We find $f_\mathrm{gas}=0.130 \pm 0.006$, consistent with $0.125 \pm 0.005$ \citep{mantz16}. The intrinsic scatter is high, but consistent within 1$\sigma$ with the $\sim$20\% found by other studies \citep{becker11, mahdavi13, farahi18}. If we remove the very disturbed clusters, for which the weak-lensing masses are most likely to be biased, the intrinsic scatter drops by 1$\sigma$ to $19 \pm 16$\% in line with expectations from \citet[e.g.][]{applegate16}. The best fit $f_\mathrm{gas}$ changes to $f_\mathrm{gas}=0.126 \pm 0.007$ for the remaining 27 clusters. For the 8 clusters in our sample which \citet{mantz15b} named relaxed, we find a value of $0.139 \pm 0.014$, consistent within the errorbars with our other estimates and the value of \citet{mantz16}.

\section{Conclusions}
\label{sec:conclusions}

Galaxy cluster counts have the potential to put tight constraints on cosmological parameters, if large numbers of clusters with accurate mass estimates are observed. 
The Multi Epoch Nearby Cluster Survey and Canadian Cluster Comparison Project provide high quality optical imaging data in the $g$ and $r$ filters observed using the Canada-France-Hawaii Telescope (CFHT) for a sample of $\sim$100 galaxy clusters. We performed a thorough weak-lensing analysis on this sample, excluding some of the clusters because of their very high Galactic extinction, which prevented us from establishing a robust correction for contamination by cluster members for those clusters. 
We used updated redshift catalogues of the COSMOS field to determine a mean lensing efficiency reliable to 9\% for the highest redshift clusters and on average accurate to $\sim$2\%. The photometric redshift distribution is one of the largest sources of error in our analysis. 
For the low redshift \mn \ clusters trading off multi-wavelength information against number of observed clusters has proven worth-while. However, precision can be increased using redshift distributions for individual galaxies \citep{applegate14} and our analysis is limited by the depth and area of the auxiliary redshift catalogues.

The radial profiles of the corrected tangential shear were fit with parametric models to estimate cluster masses, as well as used to determine aperture masses. Both methods are in agreement on the masses. We calibrate our mass modelling pipelines using the state-of-the-art HYDRANGEA numerical simulations of galaxy clusters. Both methods show only  $\lesssim$4\% percent level biases with uncertainties of 2-3\% at $R_{500}$ in the cluster simulations and we corrected for these biases. The overall average systematic uncertainty for our masses is $\lesssim$5\% similar to the statistical uncertainty.

Finally, we calculated the scaling relation between weak-lensing masses and \textit{Planck} mass estimates for 61 clusters, resulting in a bias of $1-b=0.84 \pm 0.04$. This value is somewhat higher than the estimate in \citetalias{hoekstra15}, mainly due to the use of the updated photometric redshift catalogue. The sample shows no significant trend with either mass or redshift, but simple tests show that our selection of clusters might result in a slightly higher $1-b$ up to a maximum change of 0.07. This highlights the importance of modelling the selection function for cosmological analyses. The gas fraction of clusters relates to the matter density in the Universe, and for relaxed clusters the uncertainty in this relation from baryonic processes should be small. A comparison of lensing mass and gas mass at $r_{500}$ produced a gas fraction $M_\mathrm{gas} / M_\mathrm{WL} = 0.139 \pm 0.014$ for 8 relaxed clusters. This value is consistent with the value found by \citet{mantz16}.

Weak-lensing calibration of cluster observables is the limiting factor for cluster cosmology and large weak-lensing surveys are required for this calibration. The combination of the \mn \ and CCCP surveys provides such a large sample for the some of the most massive clusters in the Universe, over a large range of redshifts and cluster masses. Future improvements of the weak-lensing analysis, in particular the photometric redshift distribution and calibration of mass modelling with simulations, will further improve our ability to constrain the scaling relations.

\section*{Acknowledgements}

We would like to thank Adam Muzzin for access to the redshift catalogue, Nick Battaglia for helpful comments and Adam Mantz for useful feedback on the scaling relation analysis.
We thank Melissa Graham for her work on the \mn \ observations and 
Claudio Dalla Vecchia for his work on the HYDRANGEA simulations.
This work is based on observations obtained with MegaPrime/MegaCam, a joint project of CFHT and CEA/IRFU, at the Canada-France-Hawaii Telescope (CFHT) which is operated by the National Research Council (NRC) of Canada, the Institut National des Science de l'Univers of the Centre National de la Recherche Scientifique (CNRS) of France, and the University of Hawaii. This work is based in part on data products produced at Terapix available at the Canadian Astronomy Data Centre as part of the Canada-France-Hawaii Telescope Legacy Survey, a collaborative project of NRC and CNRS. 

RH and AvdL are supported by the US Department of Energy under award DE-SC0018053. RH, CS, HH acknowledge support from the European Research Council FP7 grant number 279396. Research by DJS is supported by NSF grants AST-1821967, 1821987, 1813708, 1813466, and 1908972. 
YMB acknowledges funding from the EU Horizon 2020 research and innovation programme under Marie Sk{\l}odowska-Curie grant agreement 747645 (ClusterGal) and the Netherlands Organisation for Scientific Research (NWO) through VENI grant 016.183.011.

\bibliographystyle{mnras}
\bibliography{masses}

\appendix

\section{Cluster properties}

Here we document all the cluster properties calculated in this work. Table~\ref{tbl:redshifts} lists general properties of the clusters used in this work, as well as details on the observations, which were used to compute the boost correction for individual clusters. The calculated lensing efficiencies and the mass ranges used for the weak-lensing analysis are also presented. Table~\ref{tbl:masses} lists the results of the weak-lensing analysis and the state of the cluster determined using the results of \citet{mantz15b}.


\begin{table*}
\begin{tabular*}{1.0\textwidth}{ l l c c c r c c c c c c c c l l }
\multicolumn{1}{c}{(1) } & \multicolumn{1}{c}{ (2) } & \multicolumn{1}{c}{ (3) } & \multicolumn{1}{c}{ (4) } & \multicolumn{1}{c}{ (5) } &  \multicolumn{1}{c}{ (6) } & \multicolumn{1}{c}{ (7) } & \multicolumn{1}{c}{ (8) } &  \multicolumn{1}{c}{ (9) } & \multicolumn{1}{c}{ (10) } & \multicolumn{1}{c}{ (11) } & \multicolumn{1}{c}{ (12) } & \multicolumn{1}{c}{ (13) } \\

\multicolumn{1}{c}{ } & \multicolumn{1}{c}{ cluster} & \multicolumn{1}{c}{ $z$} &  \multicolumn{1}{c}{ RA$^{\mathrm{BCG}}$} & \multicolumn{1}{c}{ Dec$^{\mathrm{BCG}}$} & \multicolumn{1}{c}{ $\langle \beta \rangle$} &  \multicolumn{1}{c}{ $\langle \beta^2 \rangle$} & \multicolumn{1}{c}{ $\delta \beta$} & \multicolumn{1}{c}{ $A_r$} & \multicolumn{1}{c}{seeing}  & \multicolumn{1}{c}{ depth} & \multicolumn{1}{c}{$R_2$-$R_\mathrm{max}$} & \multicolumn{1}{c}{ $m_r$ }\\
 &  & & (J2000) & \multicolumn{1}{c}{(J2000)}  &  &  &  &  mag & arcsec & mag/arcsec$^2$ & Mpc & \\
 
 \hline
 \\
1 & A7 & 0.106 & 00:11:45.25 & +32:24:56.5 & 0.783 & 0.643 & 0.005 \hspace{8pt}& 0.09 & 0.60 & 26.53 \hspace{8pt}& 2.0--2.9 \hspace{8pt}& 20--24.5 \\
2 & A21 & 0.095 & 00:20:36.97 & +28:39:33.0 & 0.800 & 0.668 & 0.004 \hspace{8pt}& 0.08 & 0.63 & 26.49 \hspace{8pt}& 1.9--2.6 \hspace{8pt}& 20--24.5 \\
3 & A85 & 0.055 & 00:41:50.44 & \hspace{1.5pt}--\hspace{0.5pt}09:18:11.0 & 0.878 & 0.786 & 0.002 \hspace{8pt}& 0.08 & 0.62 & 26.39 \hspace{8pt}& 1.4--1.6 \hspace{8pt}& 20--24.5 \\
4 & A119 & 0.044 & 00:56:16.09 & \hspace{1.5pt}--\hspace{0.5pt}01:15:19.0 & 0.901 & 0.823 & 0.002 \hspace{8pt}& 0.08 & 0.64 & 26.43 \hspace{8pt}& 1.2--1.3 \hspace{8pt}& 20--24.5 \\
5 & A133 & 0.057 & 01:02:41.70 & \hspace{1.5pt}--\hspace{0.5pt}21:52:55.2 & 0.873 & 0.778 & 0.002 \hspace{8pt}& 0.04 & 0.68 & 26.42 \hspace{8pt}& 1.4--1.7 \hspace{8pt}& 20--24.5 \\
6 & A646 & 0.129 & 08:22:09.53 & +47:05:53.3 & 0.740 & 0.584 & 0.007 \hspace{8pt}& 0.09 & 0.68 & 26.44 \hspace{8pt}& 2.2--3.5 \hspace{8pt}& 20--24.5 \\
7 & A655 & 0.127 & 08:25:29.04 & +47:08:00.8 & 0.744 & 0.589 & 0.007 \hspace{8pt}& 0.08 & 0.65 & 26.44 \hspace{8pt}& 2.2--3.4 \hspace{8pt}& 20--24.5 \\
8 & A754 & 0.054 & 09:08:32.36 & \hspace{1.5pt}--\hspace{0.5pt}09:37:47.2 & 0.878 & 0.786 & 0.002 \hspace{8pt}& 0.15 & 0.74 & 26.49 \hspace{8pt}& 1.4--1.6 \hspace{8pt}& 20--24.5 \\
9 & A780 & 0.054 & 09:18:05.66 & \hspace{1.5pt}--\hspace{0.5pt}12:05:43.7 & 0.877 & 0.783 & 0.002 \hspace{8pt}& 0.09 & 0.80 & 26.38 \hspace{8pt}& 1.3--1.6 \hspace{8pt}& 20--24.5 \\
10 & A795 & 0.136 & 09:24:05.28 & +14:10:21.7 & 0.721 & 0.558 & 0.006 \hspace{8pt}& 0.06 & 0.72 & 26.34 \hspace{8pt}& 2.3--3.6 \hspace{8pt}& 20--24.5 \\
11 & A961 & 0.124 & 10:16:22.80 & +33:38:17.7 & 0.743 & 0.588 & 0.006 \hspace{8pt}& 0.04 & 0.71 & 26.33 \hspace{8pt}& 2.2--3.3 \hspace{8pt}& 20--24.5 \\
12 & A990 & 0.144 & 10:23:39.90 & +49:08:38.7 & 0.710 & 0.544 & 0.007 \hspace{8pt}& 0.01 & 0.78 & 26.37 \hspace{8pt}& 2.3--3.8 \hspace{8pt}& 20--24.5 \\
13 & A1033 & 0.126 & 10:31:44.32 & +35:02:29.1 & 0.740 & 0.584 & 0.006 \hspace{8pt}& 0.04 & 0.65 & 26.27 \hspace{8pt}& 2.1--3.4 \hspace{8pt}& 20--24.5 \\
14 & A1068 & 0.138 & 10:40:44.47 & +39:57:11.4 & 0.724 & 0.563 & 0.007 \hspace{8pt}& 0.05 & 0.61 & 26.30 \hspace{8pt}& 2.2--3.7 \hspace{8pt}& 20--24.5 \\
15 & A1132 & 0.136 & 10:58:23.64 & +56:47:42.0 & 0.723 & 0.561 & 0.007 \hspace{8pt}& 0.02 & 0.68 & 26.28 \hspace{8pt}& 2.3--3.6 \hspace{8pt}& 20--24.5 \\
16 & A1285 & 0.106 & 11:30:23.80 & \hspace{1.5pt}--\hspace{0.5pt}14:34:52.2 & 0.768 & 0.622 & 0.003 \hspace{8pt}& 0.09 & 0.81 & 26.29 \hspace{8pt}& 2.0--2.9 \hspace{8pt}& 20--24.5 \\
17 & A1348 & 0.119 & 11:41:24.18 & \hspace{1.5pt}--\hspace{0.5pt}12:16:38.4 & 0.747 & 0.592 & 0.004 \hspace{8pt}& 0.07 & 0.82 & 26.33 \hspace{8pt}& 2.1--3.2 \hspace{8pt}& 20--24.5 \\
18 & A1361 & 0.117 & 11:43:39.60 & +46:21:20.7 & 0.762 & 0.614 & 0.006 \hspace{8pt}& 0.05 & 0.61 & 26.43 \hspace{8pt}& 2.1--3.2 \hspace{8pt}& 20--24.5 \\
19 & A1413 & 0.143 & 11:55:18.00 & +23:24:18.1 & 0.713 & 0.548 & 0.008 \hspace{8pt}& 0.05 & 0.65 & 26.41 \hspace{8pt}& 2.3--3.8 \hspace{8pt}& 20--24.5 \\
20 & A1650 & 0.084 & 12:58:41.49 & \hspace{1.5pt}--\hspace{0.5pt}01:45:41.0 & 0.819 & 0.695 & 0.003 \hspace{8pt}& 0.04 & 0.76 & 26.50 \hspace{8pt}& 1.8--2.4 \hspace{8pt}& 20--24.5 \\
21 & A1651 & 0.085 & 12:59:22.49 & \hspace{1.5pt}--\hspace{0.5pt}04:11:45.7 & 0.807 & 0.677 & 0.003 \hspace{8pt}& 0.06 & 0.90 & 26.27 \hspace{8pt}& 1.8--2.4 \hspace{8pt}& 20--24.5 \\
22 & A1781 & 0.062 & 13:44:52.54 & +29:46:15.6 & 0.865 & 0.766 & 0.002 \hspace{8pt}& 0.04 & 0.73 & 26.60 \hspace{8pt}& 1.5--1.8 \hspace{8pt}& 20--24.5 \\
23 & A1795 & 0.062 & 13:48:52.49 & +26:35:34.8 & 0.864 & 0.764 & 0.002 \hspace{8pt}& 0.03 & 0.68 & 26.43 \hspace{8pt}& 1.5--1.8 \hspace{8pt}& 20--24.5 \\
24 & A1927 & 0.095 & 14:31:06.78 & +25:38:01.6 & 0.803 & 0.673 & 0.004 \hspace{8pt}& 0.08 & 0.62 & 26.52 \hspace{8pt}& 1.9--2.6 \hspace{8pt}& 20--24.5 \\
25 & A1991 & 0.059 & 14:54:31.48 & +18:38:33.3 & 0.869 & 0.771 & 0.002 \hspace{8pt}& 0.07 & 0.67 & 26.38 \hspace{8pt}& 1.4--1.7 \hspace{8pt}& 20--24.5 \\
26 & A2029 & 0.077 & 15:10:56.09 & +05:44:41.3 & 0.834 & 0.717 & 0.002 \hspace{8pt}& 0.08 & 0.65 & 26.48 \hspace{8pt}& 1.7--2.2 \hspace{8pt}& 20--24.5 \\
27 & A2033 & 0.082 & 15:11:26.51 & +06:20:56.7 & 0.826 & 0.706 & 0.003 \hspace{8pt}& 0.08 & 0.61 & 26.56 \hspace{8pt}& 1.8--2.3 \hspace{8pt}& 20--24.5 \\
28 & A2050 & 0.118 & 15:16:17.92 & +00:05:20.9 & 0.760 & 0.611 & 0.006 \hspace{8pt}& 0.12 & 0.62 & 26.51 \hspace{8pt}& 2.1--3.2 \hspace{8pt}& 20--24.5 \\
29 & A2055 & 0.102 & 15:18:45.70 & +06:13:56.3 & 0.788 & 0.651 & 0.004 \hspace{8pt}& 0.08 & 0.61 & 26.51 \hspace{8pt}& 2.0--2.8 \hspace{8pt}& 20--24.5 \\
30 & A2064 & 0.108 & 15:20:52.24 & +48:39:38.7 & 0.780 & 0.639 & 0.005 \hspace{8pt}& 0.04 & 0.69 & 26.66 \hspace{8pt}& 2.1--3.0 \hspace{8pt}& 20--24.5 \\
31 & A2065 & 0.073 & 15:22:29.16 & +27:42:27.7 & 0.842 & 0.730 & 0.002 \hspace{8pt}& 0.09 & 0.65 & 26.57 \hspace{8pt}& 1.7--2.1 \hspace{8pt}& 20--24.5 \\
32 & A2069 & 0.116 & 15:24:07.46 & +29:53:20.4 & 0.765 & 0.618 & 0.006 \hspace{8pt}& 0.05 & 0.61 & 26.64 \hspace{8pt}& 2.1--3.2 \hspace{8pt}& 20--24.5 \\
33 & A2142 & 0.091 & 15:58:19.98 & +27:14:00.4 & 0.809 & 0.680 & 0.003 \hspace{8pt}& 0.10 & 0.62 & 26.54 \hspace{8pt}& 1.9--2.5 \hspace{8pt}& 20--24.5 \\
34 & A2420 & 0.085 & 22:10:18.76 & \hspace{1.5pt}--\hspace{0.5pt}12:10:13.9 & 0.814 & 0.688 & 0.002 \hspace{8pt}& 0.13 & 0.67 & 26.26 \hspace{8pt}& 1.8--2.4 \hspace{8pt}& 20--24.5 \\
35 & A2426 & 0.098 & 22:14:31.57 & \hspace{1.5pt}--\hspace{0.5pt}10:22:26.1 & 0.785 & 0.645 & 0.003 \hspace{8pt}& 0.13 & 0.72 & 26.18 \hspace{8pt}& 2.0--2.7 \hspace{8pt}& 20--24.5 \\
36 & A2440 & 0.091 & 22:23:56.92 & \hspace{1.5pt}--\hspace{0.5pt}01:34:59.4 & 0.800 & 0.667 & 0.003 \hspace{8pt}& 0.17 & 0.69 & 26.39 \hspace{8pt}& 1.9--2.5 \hspace{8pt}& 20--24.5 \\
37 & A2443 & 0.108 & 22:26:07.92 & +17:21:23.7 & 0.775 & 0.632 & 0.005 \hspace{8pt}& 0.14 & 0.61 & 26.47 \hspace{8pt}& 2.1--3.0 \hspace{8pt}& 20--24.5 \\
38 & A2495 & 0.078 & 22:50:19.71 & +10:54:12.8 & 0.832 & 0.715 & 0.003 \hspace{8pt}& 0.17 & 0.61 & 26.46 \hspace{8pt}& 1.7--2.2 \hspace{8pt}& 20--24.5 \\
39 & A2597 & 0.085 & 23:25:19.72 & \hspace{1.5pt}--\hspace{0.5pt}12:07:26.6 & 0.815 & 0.689 & 0.003 \hspace{8pt}& 0.07 & 0.66 & 26.25 \hspace{8pt}& 1.8--2.4 \hspace{8pt}& 20--24.5 \\
40 & A2627 & 0.126 & 23:36:42.07 & +23:55:29.4 & 0.743 & 0.588 & 0.006 \hspace{8pt}& 0.17 & 0.64 & 26.46 \hspace{8pt}& 2.1--3.4 \hspace{8pt}& 20--24.5 \\
41 & A2670 & 0.076 & 23:54:13.67 & \hspace{1.5pt}--\hspace{0.5pt}10:25:08.1 & 0.832 & 0.714 & 0.002 \hspace{8pt}& 0.10 & 0.76 & 26.34 \hspace{8pt}& 1.7--2.2 \hspace{8pt}& 20--24.5 \\
42 & A2703 & 0.114 & 00:05:23.94 & +16:13:09.3 & 0.766 & 0.620 & 0.005 \hspace{8pt}& 0.10 & 0.59 & 26.49 \hspace{8pt}& 2.1--3.1 \hspace{8pt}& 20--24.5 \\
43 & MKW3S & 0.045 & 15:21:51.80 & +07:42:31.8 & 0.901 & 0.823 & 0.002 \hspace{8pt}& 0.08 & 0.64 & 26.55 \hspace{8pt}& 1.1--1.3 \hspace{8pt}& 20--24.5 \\
44 & RXJ0132 & 0.149 & 01:32:41.10 & \hspace{1.5pt}--\hspace{0.5pt}08:04:04.5 & 0.708 & 0.542 & 0.009 \hspace{8pt}& 0.07 & 0.60 & 26.39 \hspace{8pt}& 2.3--3.9 \hspace{8pt}& 20--24.5 \\
45 & RXJ0736 & 0.118 & 07:36:38.08 & +39:24:52.8 & 0.751 & 0.599 & 0.005 \hspace{8pt}& 0.10 & 0.69 & 26.41 \hspace{8pt}& 2.1--3.2 \hspace{8pt}& 20--24.5 \\
46 & RXJ2344 & 0.079 & 23:44:18.19 & \hspace{1.5pt}--\hspace{0.5pt}04:22:48.7 & 0.829 & 0.709 & 0.002 \hspace{8pt}& 0.08 & 0.69 & 26.34 \hspace{8pt}& 1.7--2.2 \hspace{8pt}& 20--24.5 \\
47 & ZWCL1023 & 0.143 & 10:25:57.97 & +12:41:08.7 & 0.707 & 0.541 & 0.007 \hspace{8pt}& 0.10 & 0.72 & 26.33 \hspace{8pt}& 2.3--3.8 \hspace{8pt}& 20--24.5 \\
48 & ZWCL1215 & 0.075 & 12:17:41.13 & +03:39:21.2 & 0.833 & 0.716 & 0.002 \hspace{8pt}& 0.04 & 0.86 & 26.37 \hspace{8pt}& 1.7--2.1 \hspace{8pt}& 20--24.5 \\
49 & A68 & 0.255 & 00:37:06.90 & +09:09:24.0 & 0.601 & 0.410 & 0.023 \hspace{8pt}& & & \hspace{8pt}& 2.4--3.2 \hspace{8pt}& 22--25.0 \\
50 & A209 & 0.206 & 01:31:52.50 & \hspace{1.5pt}--\hspace{0.5pt}13:36:39.9 & 0.667 & 0.488 & 0.019 \hspace{8pt}& & & \hspace{8pt}& 2.0--2.7 \hspace{8pt}& 22--25.0 \\
51 & A267 & 0.23 & 01:52:42.00 & +01:00:25.9 & 0.630 & 0.443 & 0.021 \hspace{8pt}& & & \hspace{8pt}& 2.2--2.9 \hspace{8pt}& 22--25.0 \\
52 & A370 & 0.375 & 02:39:52.69 & \hspace{1.5pt}--\hspace{0.5pt}01:34:18.0 & 0.470 & 0.279 & 0.027 \hspace{8pt}& & & \hspace{8pt}& 3.1--4.1 \hspace{8pt}& 22--25.0 \\
53 & A383 & 0.187 & 02:48:03.40 & \hspace{1.5pt}--\hspace{0.5pt}03:31:44.0 & 0.675 & 0.497 & 0.017 \hspace{8pt}& & & \hspace{8pt}& 1.9--2.5 \hspace{8pt}& 22--24.5 \\
54 & A963 & 0.206 & 10:17:03.79 & +39:02:51.0 & 0.662 & 0.483 & 0.019 \hspace{8pt}& & & \hspace{8pt}& 2.0--2.7 \hspace{8pt}& 22--25.0 \\
55 & A1689 & 0.183 & 13:11:30.00 & \hspace{1.5pt}--\hspace{0.5pt}01:20:30.0 & 0.685 & 0.510 & 0.017 \hspace{8pt}& & & \hspace{8pt}& 1.8--2.5 \hspace{8pt}& 22--24.5 \\
56 & A1763 & 0.223 & 13:35:20.10 & +41:00:03.9 & 0.633 & 0.447 & 0.020 \hspace{8pt}& & & \hspace{8pt}& 2.2--2.9 \hspace{8pt}& 22--25.0 \\
57 & A2218 & 0.176 & 16:35:48.79 & +66:12:51.0 & 0.683 & 0.506 & 0.016 \hspace{8pt}& & & \hspace{8pt}& 1.8--2.4 \hspace{8pt}& 22--24.5 \\
58 & A2219 & 0.226 & 16:40:19.90 & +46:42:41.0 & 0.640 & 0.455 & 0.021 \hspace{8pt}& & & \hspace{8pt}& 2.2--2.9 \hspace{8pt}& 22--25.0 \\
59 & A2390 & 0.228 & 21:53:36.79 & +17:41:44.0 & 0.642 & 0.458 & 0.021 \hspace{8pt}& & & \hspace{8pt}& 2.2--2.9 \hspace{8pt}& 22--25.0 \\
60 & MS0016 & 0.547 & 00:18:33.49 & +16:26:16.0 & 0.341 & 0.173 & 0.030 \hspace{8pt}& & & \hspace{8pt}& 3.8--5.1 \hspace{8pt}& 22--25.0 \\
61 & MS0906 & 0.17 & 09:09:12.60 & +10:58:28.0 & 0.716 & 0.550 & 0.016 \hspace{8pt}& & & \hspace{8pt}& 1.7--2.3 \hspace{8pt}& 22--25.0 \\
62 & MS1224 & 0.326 & 12:27:13.50 & +19:50:56.0 & 0.518 & 0.323 & 0.026 \hspace{8pt}& & & \hspace{8pt}& 2.8--3.8 \hspace{8pt}& 22--25.0 \\
63 & MS1231 & 0.235 & 12:33:55.39 & +15:25:58.0 & 0.632 & 0.446 & 0.022 \hspace{8pt}& & & \hspace{8pt}& 2.2--3.0 \hspace{8pt}& 22--25.0 \\
64 & MS1358 & 0.329 & 13:59:50.59 & +62:31:05.0 & 0.519 & 0.325 & 0.026 \hspace{8pt}& & & \hspace{8pt}& 2.8--3.8 \hspace{8pt}& 22--25.0 \\

 \end{tabular*}
\end{table*}

\begin{table*}
\begin{tabular*}{1.0\textwidth}{ l l c c c r c c c c c c c c c c }
\multicolumn{1}{c}{(1) } & \multicolumn{1}{c}{ (2) } & \multicolumn{1}{c}{ (3) } & \multicolumn{1}{c}{ (4) } & \multicolumn{1}{c}{ (5) } &  \multicolumn{1}{c}{ (6) } & \multicolumn{1}{c}{ (7) } & \multicolumn{1}{c}{ (8) } &  \multicolumn{1}{c}{ (9) } & \multicolumn{1}{c}{ (10) } & \multicolumn{1}{c}{ (11) } & \multicolumn{1}{c}{ (12) } & \multicolumn{1}{c}{ (13) }  \\

\multicolumn{1}{c}{ } & \multicolumn{1}{c}{ cluster} & \multicolumn{1}{c}{ $z$} &  \multicolumn{1}{c}{ RA$^{\mathrm{BCG}}$} & \multicolumn{1}{c}{ Dec$^{\mathrm{BCG}}$} & \multicolumn{1}{c}{ $\langle \beta \rangle$} &  \multicolumn{1}{c}{ $\langle \beta^2 \rangle$} & \multicolumn{1}{c}{ $\delta \beta$} & \multicolumn{1}{c}{ $A_r$} & \multicolumn{1}{c}{seeing}  & \multicolumn{1}{c}{ depth} & \multicolumn{1}{c}{$R_2$-$R_\mathrm{max}$} & \multicolumn{1}{c}{$m_r$} \\
 &  & & (J2000) & \multicolumn{1}{c}{(J2000)}  &  &  &  &  mag & arcsec & mag/arcsec$^2$  & Mpc & \\
 
 \hline
 \\
65 & MS1455 & 0.257 & 14:57:15.10 & +22:20:35.0 & 0.612 & 0.423 & 0.023 \hspace{8pt}& & & \hspace{8pt}& 2.4--3.2 \hspace{8pt}& 22--25.0 \\
66 & MS1512 & 0.373 & 15:14:22.50 & +36:36:20.9 & 0.484 & 0.291 & 0.028 \hspace{8pt}& & & \hspace{8pt}& 3.1--4.1 \hspace{8pt}& 22--25.0 \\
67 & MS1621 & 0.428 & 16:23:35.50 & +26:34:14.0 & 0.433 & 0.245 & 0.029 \hspace{8pt}& & & \hspace{8pt}& 3.4--4.5 \hspace{8pt}& 22--25.0 \\
68 & CL0024 & 0.39 & 00:26:35.59 & +17:09:43.9 & 0.450 & 0.261 & 0.027 \hspace{8pt}& & & \hspace{8pt}& 3.2--4.2 \hspace{8pt}& 22--25.0 \\
69 & A115N & 0.197 & 00:55:50.60 & +26:24:37.6 & 0.685 & 0.513 & 0.020 \hspace{8pt}& & & \hspace{8pt}& 2.9--4.9 \hspace{8pt}& 22--25.0 \\
70 & A115S & 0.197 & 00:56:00.25 & +26:20:32.7 & 0.685 & 0.513 & 0.020 \hspace{8pt}& & & \hspace{8pt}& 2.9--4.9 \hspace{8pt}& 22--25.0 \\
71 & A222 & 0.213 & 01:37:34.00 & \hspace{1.5pt}--\hspace{0.5pt}12:59:29.0 & 0.661 & 0.484 & 0.021 \hspace{8pt}& & & \hspace{8pt}& 3.1--5.2 \hspace{8pt}& 22--25.0 \\
72 & A223N & 0.207 & 01:38:02.20 & \hspace{1.5pt}--\hspace{0.5pt}12:45:19.6 & 0.669 & 0.494 & 0.020 \hspace{8pt}& & & \hspace{8pt}& 3.0--5.1 \hspace{8pt}& 22--25.0 \\
73 & A223S & 0.207 & 01:37:55.90 & \hspace{1.5pt}--\hspace{0.5pt}12:49:09.9 & 0.669 & 0.494 & 0.020 \hspace{8pt}& & & \hspace{8pt}& 3.0--5.1 \hspace{8pt}& 22--25.0 \\
74 & A520 & 0.199 & 04:54:19.88 & +02:57:44.6 & 0.682 & 0.509 & 0.020 \hspace{8pt}& & & \hspace{8pt}& 3.0--4.9 \hspace{8pt}& 22--25.0 \\
75 & A521 & 0.253 & 04:54:06.88 & \hspace{1.5pt}--\hspace{0.5pt}10:13:24.7 & 0.605 & 0.419 & 0.022 \hspace{8pt}& & & \hspace{8pt}& 3.5--5.9 \hspace{8pt}& 22--25.0 \\
76 & A586 & 0.171 & 07:32:20.20 & +31:38:00.7 & 0.703 & 0.535 & 0.014 \hspace{8pt}& & & \hspace{8pt}& 2.6--4.4 \hspace{8pt}& 22--25.0 \\
77 & A611 & 0.288 & 08:00:56.81 & +36:03:23.6 & 0.561 & 0.371 & 0.023 \hspace{8pt}& & & \hspace{8pt}& 3.9--6.5 \hspace{8pt}& 22--25.0 \\
78 & A697 & 0.282 & 08:42:57.55 & +36:21:59.2 & 0.580 & 0.391 & 0.025 \hspace{8pt}& & & \hspace{8pt}& 3.8--6.4 \hspace{8pt}& 22--25.0 \\
79 & A851 & 0.407 & 09:42:57.45 & +46:58:49.7 & 0.448 & 0.261 & 0.028 \hspace{8pt}& & & \hspace{8pt}& 4.9--8.1 \hspace{8pt}& 22--25.0 \\
80 & A959 & 0.286 & 10:17:34.34 & +59:33:39.0 & 0.577 & 0.388 & 0.025 \hspace{8pt}& & & \hspace{8pt}& 3.9--6.5 \hspace{8pt}& 22--25.0 \\
81 & A1234 & 0.166 & 11:22:29.92 & +21:24:21.6 & 0.723 & 0.561 & 0.016 \hspace{8pt}& & & \hspace{8pt}& 2.6--4.3 \hspace{8pt}& 22--25.0 \\
82 & A1246 & 0.19 & 11:23:58.75 & +21:28:45.2 & 0.689 & 0.518 & 0.018 \hspace{8pt}& & & \hspace{8pt}& 2.9--4.8 \hspace{8pt}& 22--25.0 \\
83 & A1758 & 0.279 & 13:32:45.24 & +50:32:35.0 & 0.587 & 0.399 & 0.025 \hspace{8pt}& & & \hspace{8pt}& 3.8--6.3 \hspace{8pt}& 22--25.0 \\
84 & A1835 & 0.2533 & 14:01:02.04 & +02:52:42.7 & 0.608 & 0.422 & 0.023 \hspace{8pt}& & & \hspace{8pt}& 3.6--5.9 \hspace{8pt}& 22--25.0 \\
85 & A1914 & 0.171 & 14:25:56.69 & +37:48:59.0 & 0.720 & 0.558 & 0.017 \hspace{8pt}& & & \hspace{8pt}& 2.6--4.4 \hspace{8pt}& 22--25.0 \\
86 & A1942 & 0.224 & 14:38:21.87 & +03:40:13.2 & 0.650 & 0.470 & 0.022 \hspace{8pt}& & & \hspace{8pt}& 3.2--5.4 \hspace{8pt}& 22--25.0 \\
87 & A2104 & 0.153 & 15:40:07.90 & \hspace{1.5pt}--\hspace{0.5pt}03:18:16.0 & 0.740 & 0.584 & 0.014 \hspace{8pt}& & & \hspace{8pt}& 2.4--4.0 \hspace{8pt}& 22--25.0 \\
88 & A2111 & 0.229 & 15:39:40.50 & +34:25:27.6 & 0.642 & 0.461 & 0.022 \hspace{8pt}& & & \hspace{8pt}& 3.3--5.5 \hspace{8pt}& 22--25.0 \\
89 & A2163 & 0.203 & 16:15:33.49 & \hspace{1.5pt}--\hspace{0.5pt}06:09:16.5 & 0.658 & 0.481 & 0.017 \hspace{8pt}& & & \hspace{8pt}& 3.0--5.0 \hspace{8pt}& 22--25.0 \\
90 & A2204 & 0.152 & 16:32:46.93 & +05:34:32.9 & 0.741 & 0.586 & 0.014 \hspace{8pt}& & & \hspace{8pt}& 2.4--4.0 \hspace{8pt}& 22--25.0 \\
91 & A2259 & 0.164 & 17:20:09.65 & +27:40:08.5 & 0.724 & 0.563 & 0.015 \hspace{8pt}& & & \hspace{8pt}& 2.5--4.2 \hspace{8pt}& 22--25.0 \\
92 & A2261 & 0.224 & 17:22:27.22 & +32:07:57.9 & 0.649 & 0.469 & 0.022 \hspace{8pt}& & & \hspace{8pt}& 3.2--5.4 \hspace{8pt}& 22--25.0 \\
93 & A2537 & 0.295 & 23:08:22.21 & \hspace{1.5pt}--\hspace{0.5pt}02:11:31.6 & 0.560 & 0.370 & 0.024 \hspace{8pt}& & & \hspace{8pt}& 4.0--6.6 \hspace{8pt}& 22--25.0 \\
94 & MS0440 & 0.19 & 04:43:09.92 & +02:10:19.1 & 0.683 & 0.510 & 0.017 \hspace{8pt}& & & \hspace{8pt}& 2.9--4.8 \hspace{8pt}& 22--25.0 \\
95 & MS0451 & 0.55 & 04:54:10.83 & \hspace{1.5pt}--\hspace{0.5pt}03:00:51.7 & 0.333 & 0.166 & 0.026 \hspace{8pt}& & & \hspace{8pt}& 5.8--9.6 \hspace{8pt}& 22--25.0 \\
96 & MS1008 & 0.301 & 10:10:32.30 & \hspace{1.5pt}--\hspace{0.5pt}12:39:52.9 & 0.555 & 0.364 & 0.025 \hspace{8pt}& & & \hspace{8pt}& 4.0--6.7 \hspace{8pt}& 22--25.0 \\
97 & RXJ1347 & 0.451 & 13:47:31.85 & \hspace{1.5pt}--\hspace{0.5pt}11:45:11.1 & 0.404 & 0.223 & 0.027 \hspace{8pt}& & & \hspace{8pt}& 5.2--8.7 \hspace{8pt}& 22--25.0 \\
98 & RXJ1524 & 0.516 & 15:24:44.56 & +09:57:57.0 & 0.355 & 0.183 & 0.026 \hspace{8pt}& & & \hspace{8pt}& 5.6--9.3 \hspace{8pt}& 22--25.0 \\
99 & MACS0717 & 0.548 & 07:17:35.64 & +37:45:17.3 & 0.327 & 0.162 & 0.025 \hspace{8pt}& & & \hspace{8pt}& 5.8--9.6 \hspace{8pt}& 22--25.0 \\
100 & 3C295 & 0.46 & 14:11:20.59 & +52:12:10.0 & 0.401 & 0.220 & 0.027 \hspace{8pt}& & & \hspace{8pt}& 5.2--8.7 \hspace{8pt}& 22--25.0 \\
\hline 
 \end{tabular*}
\caption{Basic information on the clusters used in this work, parameters governing the quality of our observations, the lensing efficiency $\beta$ computed in Section~\ref{sec:photoz} and apertures for the mass analysis in Section~\ref{sec:apmass}. Columns 9, 10 and 11 are only shown for the \mn \ clusters for which they were used (see Section~\ref{sec:contamination}. (2) cluster name; (3) cluster redshift; (4)\&(5) right ascension and declination (J2000) of the BCG, which is taken to be the cluster centre; (6) the average lensing efficiency $\beta$ used to estimate the critical surface density; (7) the average $\beta^2$ used to correct the shear for the lack of individual source redshifts; (8) the error on the average lensing efficiency; (9) Galactic extinction in $r$-band magnitude; (10) seeing of the observations; (11) 1$\sigma$ depth of the observations; (12) range of the annulus used for the aperture mass measurements; (13) magnitude range in $r$-band of galaxies used for the weak-lensing analysis.  } \label{tbl:redshifts}
\end{table*}


\begin{table*}
\begin{tabularx}{1.0\textwidth}{ l l r r r r r r}
\multicolumn{1}{c}{(1) } & \multicolumn{1}{c}{ (2) } & \multicolumn{1}{c}{ (3) } & \multicolumn{1}{c}{ (4) } & \multicolumn{1}{c}{ (5) } & \multicolumn{1}{c}{ (6) } & \multicolumn{1}{c}{ (7) } & \multicolumn{1}{c}{ (8) } \\
    & \multicolumn{1}{c}{cluster}  & \multicolumn{1}{c}{$z$} &  \multicolumn{1}{c}{ $M^\mathrm{NFW}_{200}$  } & \multicolumn{1}{c}{$M^\mathrm{NFW}_{500}$} & \multicolumn{1}{c}{ $R^\mathrm{ap}_{500}$} & \multicolumn{1}{c}{$M^\mathrm{ap}_{500}$} & state \\
    
  & & & \multicolumn{1}{c}{$10^{14} M_\odot$ } &  \multicolumn{1}{c}{$10^{14} M_\odot$ } &  \multicolumn{1}{c}{ Mpc } & \multicolumn{1}{c}{$10^{14} M_\odot$ } &  \\
 
 \hline
 \\
 
1 & A7 & 0.106 & $4.4 \pm 2.3$ & $3.0 \pm 1.6$ & $0.91_{-0.06}^{+0.11}$ & $2.4_{-0.9}^{+1.3}$ & -- \\
2 & A21 & 0.095 & $6.1 \pm 2.5$ & $4.2 \pm 1.7$ & $1.17_{-0.08}^{+0.08}$ & $5.0_{-1.3}^{+1.5}$ & -- \\
3 & A85 & 0.055 & $8.4 \pm 3.3$ & $5.7 \pm 2.2$ & $1.35_{-0.14}^{+0.14}$ & $7.4_{-2.3}^{+2.3}$ & D \\
4 & A119 & 0.044 & $7.8 \pm 3.2$ & $5.3 \pm 2.2$ & $1.09_{-0.15}^{+0.15}$ & $3.9_{-1.6}^{+1.6}$ &  \\
5 & A133 & 0.057 & $4.1 \pm 2.7$ & $2.8 \pm 1.9$ & $0.92_{-0.15}^{+0.11}$ & $2.3_{-1.3}^{+1.3}$ & R \\
6 & A646 & 0.129 & $3.8 \pm 2.4$ & $2.6 \pm 1.6$ & $1.03_{-0.05}^{+0.28}$ & $3.5_{-0.9}^{+3.9}$ & -- \\
7 & A655 & 0.127 & $5.9 \pm 2.4$ & $4.0 \pm 1.7$ & $1.29_{-0.11}^{+0.16}$ & $6.8_{-1.9}^{+3.0}$ & -- \\
8 & A754 & 0.054 & $14.9 \pm 3.8$ & $10.0 \pm 2.6$ & $1.30_{-0.12}^{+0.17}$ & $6.6_{-1.9}^{+3.2}$ &  \\
9 & A780 & 0.054 & $6.5 \pm 3.0$ & $4.4 \pm 2.1$ & $1.08_{-0.14}^{+0.14}$ & $3.8_{-1.5}^{+1.5}$ & R \\
10 & A795 & 0.136 & $16.0 \pm 4.1$ & $10.6 \pm 2.8$ & $1.55_{-0.13}^{+0.11}$ & $12.2_{-3.0}^{+3.0}$ & D \\
11 & A961 & 0.124 & $7.0 \pm 2.6$ & $4.8 \pm 1.8$ & $1.18_{-0.09}^{+0.07}$ & $5.3_{-1.4}^{+1.4}$ & -- \\
12 & A990 & 0.144 & $14.1 \pm 3.4$ & $9.4 \pm 2.3$ & $1.26_{-0.10}^{+0.09}$ & $6.6_{-1.7}^{+1.8}$ & -- \\
13 & A1033 & 0.126 & $8.5 \pm 3.6$ & $5.7 \pm 2.5$ & $0.90_{-0.08}^{+0.10}$ & $2.4_{-0.9}^{+1.1}$ & -- \\
14 & A1068 & 0.138 & $5.0 \pm 2.5$ & $3.4 \pm 1.7$ & $0.88_{-0.09}^{+0.09}$ & $2.2_{-0.9}^{+0.9}$ & R \\
15 & A1132 & 0.136 & $11.2 \pm 2.7$ & $7.5 \pm 1.8$ & $1.38_{-0.06}^{+0.06}$ & $8.4_{-1.6}^{+1.5}$ &  \\
16 & A1285 & 0.106 & $6.9 \pm 2.6$ & $4.7 \pm 1.8$ & $1.16_{-0.08}^{+0.11}$ & $4.9_{-1.3}^{+1.8}$ & -- \\
17 & A1348 & 0.119 & $3.1 \pm 2.1$ & $2.1 \pm 1.5$ & $0.84_{-0.12}^{+0.17}$ & $1.9_{-1.0}^{+1.6}$ & -- \\
18 & A1361 & 0.117 & $4.8 \pm 2.4$ & $3.3 \pm 1.6$ & $0.88_{-0.12}^{+0.12}$ & $2.2_{-1.0}^{+1.2}$ & -- \\
19 & A1413 & 0.143 & $10.8 \pm 3.1$ & $7.2 \pm 2.1$ & $1.19_{-0.06}^{+0.10}$ & $5.5_{-1.1}^{+1.7}$ &  \\
20 & A1650 & 0.084 & $10.5 \pm 2.9$ & $7.1 \pm 2.0$ & $1.20_{-0.08}^{+0.07}$ & $5.3_{-1.4}^{+1.4}$ &  \\
21 & A1651 & 0.085 & $8.3 \pm 3.5$ & $5.6 \pm 2.4$ & $1.30_{-0.14}^{+0.14}$ & $6.7_{-2.2}^{+2.6}$ &  \\
22 & A1781 & 0.062 & $1.5 \pm 1.5$ & $1.1 \pm 1.1$ & $0.62_{-0.17}^{+0.27}$ & $0.7_{-0.7}^{+1.5}$ & -- \\
23 & A1795 & 0.062 & $13.9 \pm 3.3$ & $9.3 \pm 2.2$ & $1.45_{-0.11}^{+0.11}$ & $9.2_{-2.3}^{+2.3}$ &  \\
24 & A1927 & 0.095 & $4.4 \pm 2.3$ & $3.0 \pm 1.6$ & $1.03_{-0.06}^{+0.07}$ & $3.4_{-1.1}^{+1.2}$ & -- \\
25 & A1991 & 0.059 & $3.7 \pm 2.9$ & $2.5 \pm 2.0$ & $0.94_{-0.14}^{+0.21}$ & $2.5_{-1.2}^{+2.2}$ & -- \\
26 & A2029 & 0.077 & $18.1 \pm 3.8$ & $12.1 \pm 2.5$ & $1.52_{-0.10}^{+0.07}$ & $10.8_{-2.3}^{+1.9}$ & R \\
27 & A2033 & 0.082 & $3.2 \pm 2.4$ & $2.2 \pm 1.6$ & $0.80_{-0.08}^{+0.09}$ & $1.6_{-0.9}^{+0.9}$ & -- \\
28 & A2050 & 0.118 & $4.6 \pm 2.3$ & $3.1 \pm 1.6$ & $1.08_{-0.14}^{+0.07}$ & $4.0_{-1.6}^{+1.2}$ & -- \\
29 & A2055 & 0.102 & $2.9 \pm 2.2$ & $2.0 \pm 1.6$ & $0.81_{-0.09}^{+0.11}$ & $1.7_{-0.9}^{+1.0}$ & -- \\
30 & A2064 & 0.108 & $2.5 \pm 2.4$ & $1.7 \pm 1.7$ & $0.89_{-0.04}^{+0.09}$ & $2.2_{-0.8}^{+1.0}$ & -- \\
31 & A2065 & 0.073 & $12.0 \pm 3.1$ & $8.0 \pm 2.1$ & $1.37_{-0.05}^{+0.08}$ & $7.9_{-1.4}^{+1.7}$ & D \\
32 & A2069 & 0.116 & $3.2 \pm 2.2$ & $2.2 \pm 1.6$ & $0.82_{-0.06}^{+0.10}$ & $1.7_{-0.7}^{+1.0}$ & -- \\
33 & A2142 & 0.091 & $14.5 \pm 3.4$ & $9.7 \pm 2.3$ & $1.54_{-0.09}^{+0.07}$ & $11.3_{-2.2}^{+2.1}$ & D \\
34 & A2420 & 0.085 & $8.4 \pm 3.0$ & $5.7 \pm 2.0$ & $1.24_{-0.07}^{+0.09}$ & $5.9_{-1.4}^{+1.8}$ & D \\
35 & A2426 & 0.098 & $6.8 \pm 2.8$ & $4.6 \pm 1.9$ & $1.02_{-0.05}^{+0.07}$ & $3.3_{-1.0}^{+1.2}$ & D \\
36 & A2440 & 0.091 & $9.8 \pm 3.2$ & $6.6 \pm 2.2$ & $1.21_{-0.05}^{+0.06}$ & $5.5_{-1.2}^{+1.3}$ & -- \\
37 & A2443 & 0.108 & $13.4 \pm 3.3$ & $9.0 \pm 2.2$ & $1.29_{-0.06}^{+0.14}$ & $6.8_{-1.4}^{+2.7}$ & -- \\
38 & A2495 & 0.078 & $2.1 \pm 2.1$ & $1.5 \pm 1.5$ & $0.83_{-0.16}^{+0.07}$ & $1.7_{-1.1}^{+0.9}$ & -- \\
39 & A2597 & 0.085 & $3.9 \pm 2.4$ & $2.7 \pm 1.7$ & $0.95_{-0.12}^{+0.15}$ & $2.6_{-1.2}^{+1.7}$ & R \\
40 & A2627 & 0.126 & $3.0 \pm 2.3$ & $2.1 \pm 1.6$ & $0.91_{-0.22}^{+0.11}$ & $2.4_{-1.5}^{+1.2}$ & -- \\
41 & A2670 & 0.076 & $8.8 \pm 3.2$ & $5.9 \pm 2.2$ & $1.32_{-0.21}^{+0.14}$ & $7.0_{-3.0}^{+2.6}$ & -- \\
42 & A2703 & 0.114 & $5.3 \pm 2.5$ & $3.6 \pm 1.7$ & $0.88_{-0.16}^{+0.20}$ & $2.2_{-1.2}^{+2.0}$ & -- \\
43 & MKW3S & 0.045 & $2.5 \pm 2.5$ & $1.8 \pm 1.8$ & $0.82_{-0.10}^{+0.32}$ & $1.6_{-0.9}^{+2.8}$ & -- \\
44 & RXJ0132 & 0.149 & $2.6 \pm 2.1$ & $1.8 \pm 1.5$ & $0.64_{-0.04}^{+0.18}$ & $0.9_{-0.4}^{+1.0}$ & -- \\
45 & RXJ0736 & 0.118 & $2.3 \pm 2.3$ & $1.6 \pm 1.6$ & $0.57_{-0.06}^{+0.11}$ & $0.6_{-0.4}^{+0.6}$ & -- \\
46 & RXJ2344 & 0.079 & $3.8 \pm 2.7$ & $2.6 \pm 1.9$ & $0.89_{-0.08}^{+0.14}$ & $2.1_{-1.0}^{+1.4}$ &  \\
47 & ZWCL1023 & 0.143 & $4.7 \pm 2.3$ & $3.2 \pm 1.6$ & $1.06_{-0.12}^{+0.12}$ & $3.9_{-1.0}^{+1.6}$ & -- \\
48 & ZWCL1215 & 0.075 & $5.1 \pm 3.1$ & $3.5 \pm 2.2$ & $1.32_{-0.06}^{+0.12}$ & $7.0_{-1.4}^{+2.4}$ &  \\
49 & A68 & 0.255 & $11.0 \pm 2.6$ & $7.4 \pm 1.7$ & $1.35_{-0.10}^{+0.08}$ & $9.0_{-2.2}^{+2.1}$ &  \\
50 & A209 & 0.206 & $8.5 \pm 2.4$ & $5.7 \pm 1.7$ & $1.28_{-0.06}^{+0.09}$ & $7.3_{-1.5}^{+2.0}$ & D \\
51 & A267 & 0.230 & $6.4 \pm 2.5$ & $4.3 \pm 1.7$ & $1.18_{-0.07}^{+0.14}$ & $5.9_{-1.4}^{+2.6}$ & D \\
52 & A370 & 0.375 & $24.1 \pm 5.7$ & $15.7 \pm 3.7$ & $1.61_{-0.07}^{+0.07}$ & $17.5_{-2.8}^{+2.9}$ & -- \\
53 & A383 & 0.187 & $4.6 \pm 2.6$ & $3.1 \pm 1.8$ & $1.12_{-0.21}^{+0.07}$ & $4.9_{-2.4}^{+1.3}$ & R \\
54 & A963 & 0.206 & $10.3 \pm 2.6$ & $6.9 \pm 1.8$ & $1.16_{-0.10}^{+0.11}$ & $5.4_{-1.6}^{+1.9}$ &  \\
55 & A1689 & 0.183 & $23.9 \pm 4.4$ & $15.7 \pm 2.9$ & $1.53_{-0.07}^{+0.07}$ & $12.3_{-2.2}^{+2.3}$ &  \\
56 & A1763 & 0.223 & $14.1 \pm 3.8$ & $9.4 \pm 2.5$ & $1.44_{-0.06}^{+0.11}$ & $10.6_{-1.9}^{+3.0}$ & D \\
57 & A2218 & 0.176 & $13.3 \pm 4.0$ & $8.9 \pm 2.7$ & $1.28_{-0.10}^{+0.15}$ & $7.1_{-1.8}^{+2.9}$ &  \\
58 & A2219 & 0.226 & $8.2 \pm 2.1$ & $5.5 \pm 1.4$ & $1.40_{-0.07}^{+0.07}$ & $9.8_{-1.9}^{+2.0}$ &  \\
59 & A2390 & 0.228 & $16.4 \pm 3.0$ & $10.9 \pm 2.0$ & $1.27_{-0.09}^{+0.08}$ & $7.3_{-1.7}^{+1.8}$ &  \\

 \end{tabularx}
\end{table*}

\begin{table*}
\centering
\begin{tabularx}{1.0\textwidth}{ l l r r r r r r}
\multicolumn{1}{c}{(1) } & \multicolumn{1}{c}{ (2) } & \multicolumn{1}{c}{ (3) } & \multicolumn{1}{c}{ (4) } & \multicolumn{1}{c}{ (5) } & \multicolumn{1}{c}{ (6) } & \multicolumn{1}{c}{ (7) } & \multicolumn{1}{c}{ (8) } \\
    & \multicolumn{1}{c}{cluster}  & \multicolumn{1}{c}{$z$} &  \multicolumn{1}{c}{ $M^\mathrm{NFW}_{200}$  } & \multicolumn{1}{c}{$M^\mathrm{NFW}_{500}$} & \multicolumn{1}{c}{ $R^\mathrm{ap}_{500}$} & \multicolumn{1}{c}{$M^\mathrm{ap}_{500}$} & state \\
    
  & & & \multicolumn{1}{c}{$10^{14} M_\odot$ } &  \multicolumn{1}{c}{$10^{14} M_\odot$ } &  \multicolumn{1}{c}{ Mpc } & \multicolumn{1}{c}{$10^{14} M_\odot$ } &  \\
 
 \hline
 \\
 
60 & MS0016 & 0.547 & $23.4 \pm 7.4$ & $15.2 \pm 4.9$ & $1.52_{-0.09}^{+0.07}$ & $18.0_{-3.4}^{+3.1}$ &  \\
61 & MS0906 & 0.170 & $10.0 \pm 2.4$ & $6.7 \pm 1.6$ & $1.35_{-0.10}^{+0.10}$ & $8.3_{-2.1}^{+2.4}$ & D \\
62 & MS1224 & 0.326 & $4.4 \pm 2.3$ & $3.0 \pm 1.6$ & $0.87_{-0.12}^{+0.10}$ & $2.6_{-1.1}^{+1.2}$ & -- \\
63 & MS1231 & 0.235 & $1.2 \pm 1.2$ & $0.9 \pm 0.9$ & $0.53_{-0.07}^{+0.12}$ & $0.5_{-0.3}^{+0.5}$ & -- \\
64 & MS1358 & 0.329 & $11.4 \pm 2.6$ & $7.6 \pm 1.8$ & $1.28_{-0.10}^{+0.08}$ & $8.4_{-2.2}^{+2.1}$ &  \\
65 & MS1455 & 0.257 & $10.1 \pm 2.1$ & $6.8 \pm 1.4$ & $1.11_{-0.05}^{+0.08}$ & $5.0_{-1.1}^{+1.4}$ & R \\
66 & MS1512 & 0.373 & $3.8 \pm 2.5$ & $2.6 \pm 1.7$ & $0.78_{-0.09}^{+0.26}$ & $2.0_{-0.8}^{+2.8}$ & -- \\
67 & MS1621 & 0.428 & $14.8 \pm 3.6$ & $9.7 \pm 2.4$ & $1.27_{-0.06}^{+0.06}$ & $9.1_{-1.7}^{+1.9}$ & -- \\
68 & CL0024 & 0.390 & $19.2 \pm 5.3$ & $12.6 \pm 3.5$ & $1.40_{-0.07}^{+0.06}$ & $11.7_{-2.2}^{+2.1}$ & -- \\
69 & A115N & 0.197 & $4.3 \pm 2.4$ & $3.0 \pm 1.7$ & $1.07_{-0.16}^{+0.09}$ & $4.2_{-1.8}^{+1.4}$ & D \\
70 & A115S & 0.197 & $5.4 \pm 2.5$ & $3.7 \pm 1.7$ & $1.15_{-0.09}^{+0.12}$ & $5.2_{-1.4}^{+2.0}$ & D \\
71 & A222 & 0.213 & $7.0 \pm 2.5$ & $4.7 \pm 1.7$ & $1.13_{-0.04}^{+0.11}$ & $5.1_{-1.1}^{+1.9}$ & D \\
72 & A223N & 0.207 & $8.3 \pm 2.9$ & $5.6 \pm 2.0$ & $1.19_{-0.05}^{+0.15}$ & $5.9_{-1.2}^{+2.8}$ & D \\
73 & A223S & 0.207 & $6.4 \pm 2.7$ & $4.3 \pm 1.9$ & $1.34_{-0.07}^{+0.09}$ & $8.3_{-1.8}^{+2.2}$ & D \\
74 & A520 & 0.199 & $11.5 \pm 2.6$ & $7.7 \pm 1.7$ & $1.15_{-0.06}^{+0.06}$ & $5.2_{-1.2}^{+1.3}$ & D \\
75 & A521 & 0.253 & $9.1 \pm 4.0$ & $6.1 \pm 2.7$ & $1.28_{-0.07}^{+0.09}$ & $7.8_{-1.7}^{+2.1}$ & D \\
76 & A586 & 0.171 & $4.0 \pm 2.3$ & $2.7 \pm 1.6$ & $1.26_{-0.25}^{+0.09}$ & $6.7_{-3.4}^{+1.8}$ &  \\
77 & A611 & 0.288 & $7.2 \pm 2.4$ & $4.8 \pm 1.7$ & $1.18_{-0.06}^{+0.06}$ & $6.2_{-1.3}^{+1.4}$ &  \\
78 & A697 & 0.282 & $11.1 \pm 3.9$ & $7.4 \pm 2.6$ & $1.41_{-0.08}^{+0.04}$ & $10.7_{-2.1}^{+1.6}$ &  \\
79 & A851 & 0.407 & $18.8 \pm 3.7$ & $12.3 \pm 2.5$ & $1.32_{-0.08}^{+0.07}$ & $10.0_{-2.1}^{+2.1}$ & -- \\
80 & A959 & 0.286 & $16.5 \pm 4.0$ & $10.9 \pm 2.6$ & $1.35_{-0.04}^{+0.04}$ & $9.4_{-1.5}^{+1.5}$ & -- \\
81 & A1234 & 0.166 & $6.9 \pm 2.6$ & $4.6 \pm 1.8$ & $1.02_{-0.06}^{+0.06}$ & $3.5_{-1.0}^{+1.0}$ & D \\
82 & A1246 & 0.190 & $6.1 \pm 2.5$ & $4.1 \pm 1.7$ & $1.05_{-0.04}^{+0.08}$ & $4.0_{-0.9}^{+1.3}$ & D \\
83 & A1758 & 0.279 & $15.0 \pm 3.1$ & $10.0 \pm 2.1$ & $1.46_{-0.08}^{+0.06}$ & $11.8_{-2.2}^{+2.1}$ & D \\
84 & A1835 & 0.253 & $15.8 \pm 4.1$ & $10.5 \pm 2.7$ & $1.38_{-0.04}^{+0.06}$ & $9.6_{-1.5}^{+1.7}$ & R \\
85 & A1914 & 0.171 & $11.3 \pm 2.8$ & $7.6 \pm 1.9$ & $1.26_{-0.06}^{+0.08}$ & $6.8_{-1.4}^{+1.8}$ & D \\
86 & A1942 & 0.224 & $10.7 \pm 2.9$ & $7.2 \pm 2.0$ & $1.25_{-0.05}^{+0.05}$ & $6.9_{-1.3}^{+1.4}$ & -- \\
87 & A2104 & 0.153 & $14.5 \pm 3.5$ & $9.6 \pm 2.3$ & $1.43_{-0.07}^{+0.09}$ & $9.6_{-1.9}^{+2.4}$ &  \\
88 & A2111 & 0.229 & $7.9 \pm 2.7$ & $5.3 \pm 1.9$ & $1.10_{-0.09}^{+0.07}$ & $4.8_{-1.3}^{+1.3}$ & D \\
89 & A2163 & 0.203 & $13.1 \pm 3.4$ & $8.7 \pm 2.3$ & $1.42_{-0.10}^{+0.11}$ & $10.0_{-2.3}^{+2.8}$ & D \\
90 & A2204 & 0.152 & $16.8 \pm 3.3$ & $11.2 \pm 2.2$ & $1.47_{-0.05}^{+0.05}$ & $10.5_{-1.7}^{+1.8}$ & R \\
91 & A2259 & 0.164 & $6.7 \pm 2.4$ & $4.5 \pm 1.6$ & $1.12_{-0.15}^{+0.09}$ & $4.8_{-1.9}^{+1.5}$ &  \\
92 & A2261 & 0.224 & $18.7 \pm 4.3$ & $12.4 \pm 2.8$ & $1.69_{-0.07}^{+0.05}$ & $17.3_{-2.6}^{+2.3}$ &  \\
93 & A2537 & 0.295 & $16.7 \pm 3.5$ & $11.0 \pm 2.3$ & $1.31_{-0.05}^{+0.06}$ & $8.7_{-1.5}^{+1.7}$ &  \\
94 & MS0440 & 0.190 & $3.1 \pm 2.3$ & $2.1 \pm 1.6$ & $0.85_{-0.06}^{+0.06}$ & $2.1_{-0.7}^{+0.7}$ & -- \\
95 & MS0451 & 0.550 & $16.3 \pm 4.5$ & $10.6 \pm 3.0$ & $1.07_{-0.08}^{+0.06}$ & $6.3_{-1.7}^{+1.6}$ & D \\
96 & MS1008 & 0.301 & $12.0 \pm 3.0$ & $8.0 \pm 2.0$ & $1.19_{-0.06}^{+0.05}$ & $6.4_{-1.3}^{+1.3}$ & -- \\
97 & RXJ1347 & 0.451 & $16.1 \pm 5.7$ & $10.6 \pm 3.8$ & $1.32_{-0.14}^{+0.08}$ & $10.5_{-3.2}^{+2.5}$ & R \\
98 & RXJ1524 & 0.516 & $4.3 \pm 4.3$ & $2.9 \pm 2.9$ & $0.98_{-0.14}^{+0.14}$ & $4.6_{-2.4}^{+2.4}$ & D \\
99 & MACS0717 & 0.548 & $28.8 \pm 8.1$ & $18.6 \pm 5.3$ & $1.42_{-0.06}^{+0.07}$ & $14.8_{-2.6}^{+3.0}$ & D \\
100 & 3C295 & 0.460 & $8.7 \pm 3.8$ & $5.7 \pm 2.6$ & $1.12_{-0.10}^{+0.09}$ & $6.6_{-1.8}^{+1.9}$ & R \\

\hline 
 \end{tabularx}
\caption{Physical properties measured from the weak-lensing signal of the \mn \ clusters. (2) cluster name; (3) cluster redshift; (4) \& (5) Mass enclosed in a sphere where the cluster is overdense by a factor $\Delta$ compared to the critical density of the Universe, measured using NFW fitting; (6) Radius of a sphere which encloses a region overdense by a factor 500 compared to the critical density of the Universe, measured using the deprojected aperture mass method; (7) Deprojected aperture mass within $R_{500}^\mathrm{ap}$. The radius $R_{500}$ and all masses scale as $h_{70}^{-1}$; (8) State of the cluster as determined from X-ray observations. R indicates a relaxed cluster and D a very disturbed cluster. A dash indicates that there was no data for the state of the cluster in \citet{mantz15b}.} \label{tbl:masses}
\end{table*}

\section{Blank field counts}
\label{app:blanks}

The lack of deep multi-band data for galaxies in the \mn \ observations prevents us from identifying cluster members. As they are unlensed and have random orientations, these galaxies will dilute the shear signal and thus need to be corrected for. We correct for this contamination using a boost correction, for which we need to model the excess of galaxies in the cluster compared to the field as a function of cluster-centric radius. 
This approach was also used by \citetalias{hoekstra15}, who used the observations beyond 4 Mpc from the cluster center to estimate the background level for the targets observed with MegaCam. However, the MENeaCS clusters are at lower redshift, and for many targets the data do not extend that far out. Therefore, we follow a different approach to determine the expected background .

\begin{figure}
 \centering
 \includegraphics[width=8.5cm,height=8.5cm,keepaspectratio=true]{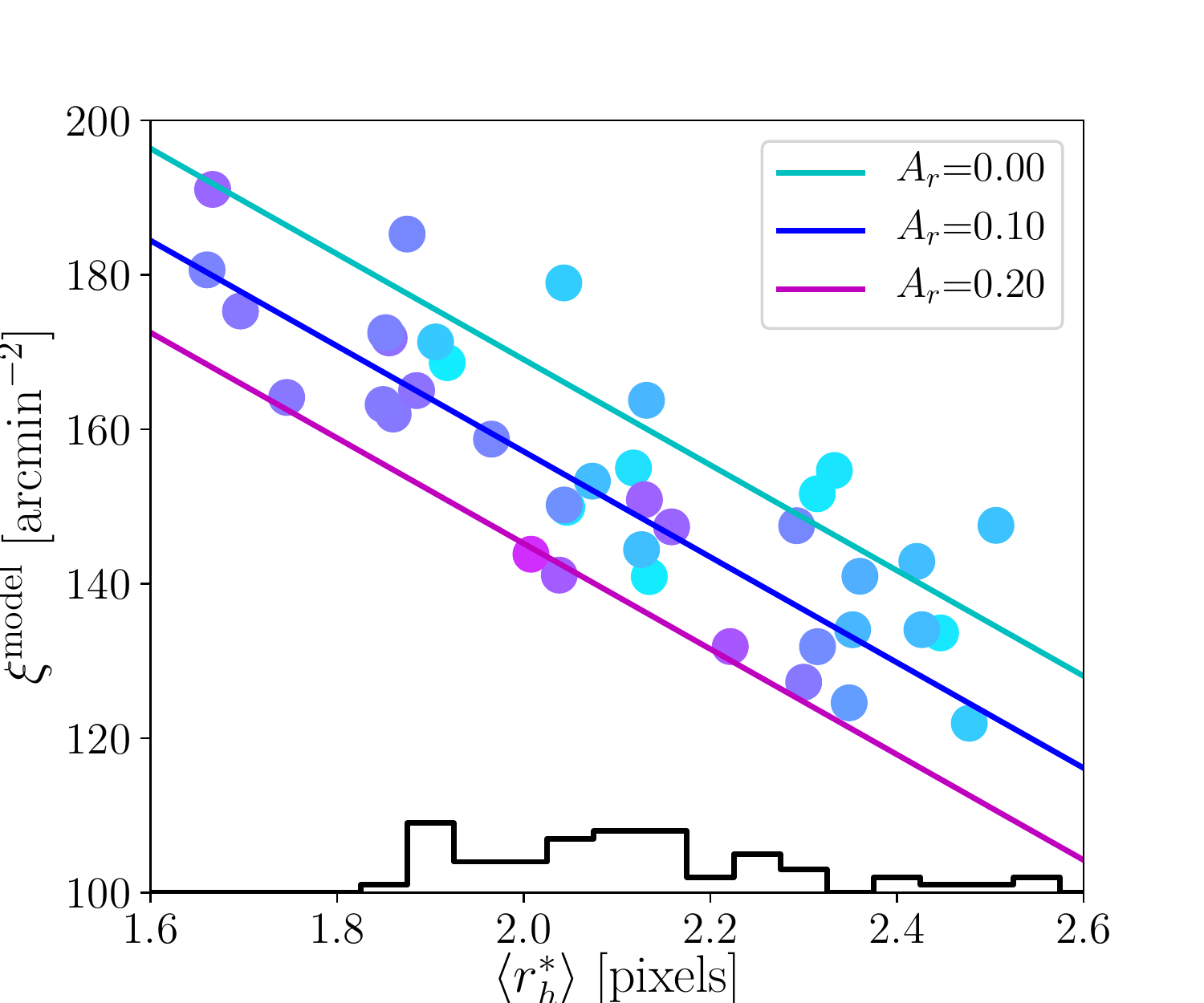}
 \caption{Weight density $\xi$ as a function of the average PSF size in the image, colour coded by Galactic extinction for 41 CFHT observations homogenised to a noise rms of 1.4 counts/pixel. Coloured lines show the best fit to the data at three levels of extinction, for the same color code as the circles. The black histogram shows the distribution of PSF sizes in the \mn \ data, covering the same range as the blank fields.}
 \label{fig:blankfit}
\end{figure}

\begin{figure*}
 \centering
 \includegraphics[width=1.0 \textwidth, trim={0 1.5cm 0 0}]{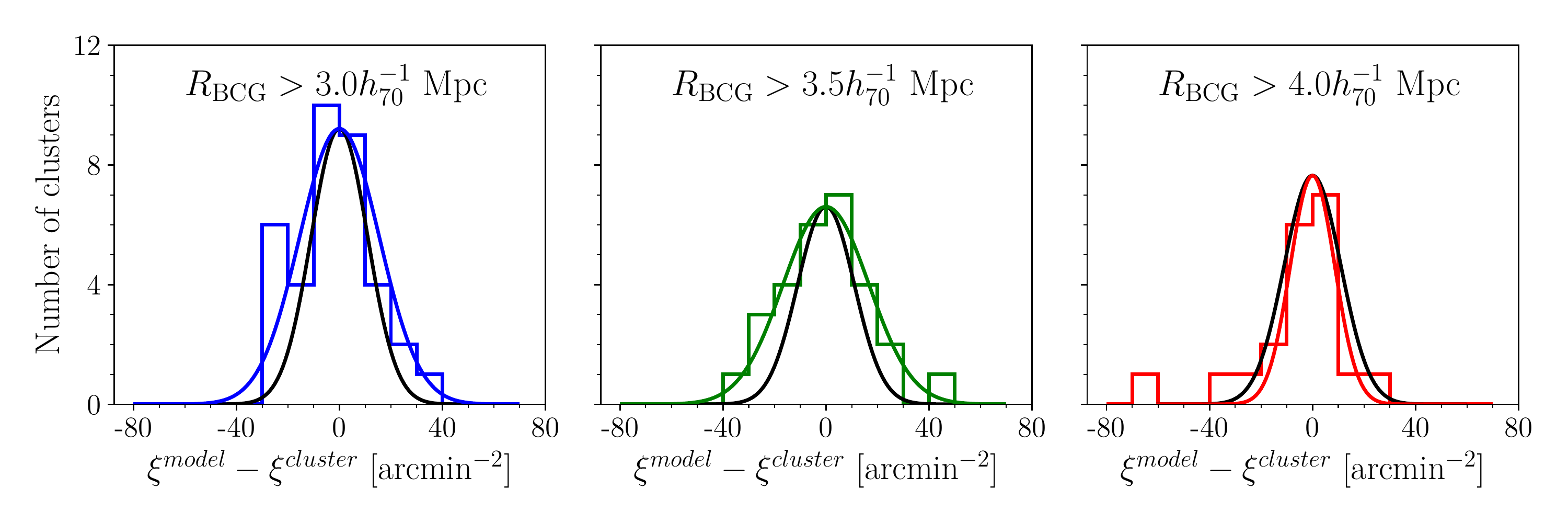} \caption{Histogram of the difference between the predicted weight density from the blank fields $\xi^{model}$ and the weight density measured outside the cluster $\xi^\mathrm{cluster}$ for the highest redshift \mn \ clusters. As the full extent of the cluster is unknown we show $\xi^\mathrm{cluster}$ for different areas corresponding to a radius of 3 Mpc, 3.5 Mpc and 4 Mpc from the BCG, from left to right respectively. The decreasing number of clusters shows that only the highest redshift clusters have any area available at large radii and highlights the need for the model prediction $\xi^\mathrm{model}$. The black curve has the same width for all three panels and it shows the distribution expected from the 6\% scatter around the best fit for the blank fields. }
 \label{fig:blankcomp}
\end{figure*}

We searched the CFHT MegaPipe \citep{Gwyn08} archive for co-added $r$-band data with a total integration $T_{\rm exp}>3600$s and image quality better than $0\farcs9$ of pointings that appeared to be blank fields (i.e. not targeting clusters).  This query resulted in 46 suitable unique targets, but upon closer inspection 
five fields had to be rejected because of a high noise level or because the Galactic extinction was too high. The remaining 41 fields all have an $r$-band Galactic extinction $A_{\rm r}$ less than 0.2 magnitude and were visually inspected to mask out obvious artifacts, leaving an effective area of approximately 33 square degrees.

The depth varies between the images and do not match the \mn \ data (\mn \ images are typically shallower). To homogenise the auxiliary data we added Gaussian noise so that all the blank fields have the same depth. We considered four r.m.s. values $\sigma_\mathrm{noise}= (1.2, 1.4, 1.6, 1.8)$ counts/pixel, corresponding to 1$\sigma$ depths of 26.60, 26.43, 26.28, 26.16 mag/arcsec$^2$, respectively, in a circular aperture of 2 pixel=0.370$''$ radius at a magnitude zeropoint of 30. This covers a range that matches most of the \mn \ data. Of the blank fields, ten and two fields are shallower than 26.60 and 26.43, respectively, so they were omitted from the sample for the analysis. The remaining homogenised images were analysed in exactly the same way as the \mn \ data, resulting in catalogs with shape measurements and corresponding uncertainties. To quantify image quality we use the half-light radius of the PSF, $r^*_{\rm h}$, where $r^*_{\rm h}=2$ pixels corresponds to a seeing of 0.63$''$. 

Figure~\ref{fig:blankfit} shows the resulting weight density $\xi$ for galaxies with $20<m_r<24.5$ as a function of the average PSF half-light radius in the image for a noise level of 1.4 counts/pixel. Using the blanks at all 4 different noise levels we fit a model to the measurements and find 
\begin{equation}
\xi^\mathrm{model} = -40.6 \sigma_\mathrm{noise} -68.4  \langle r^*_{\rm h} \rangle -122.8 A_{\rm r} + 364.2 .
\label{eq:blank_fit}
\end{equation}
The color of the circles shows the extinction for each blank field and the lines show the prediction from the fit for different extinction levels in the same color scheme. For reference we also plot the distribution of PSF half-light radii of the \mn \ observations as a black histogram in the bottom of the plot.
For the full sample we find that the r.m.s. variation in the mean weight density is 6.4\%, which is smaller than the typical statistical uncertainty in the lensing signal for an individual cluster. Hence observing clusters with a single band and modelling the excess weight as a function of cluster-centric distance is an efficient way to correct for contamination for a large sample of clusters.

The blank fields have been observed with different dither patterns than the MENeaCS data. Consequently, the variations in depth will not exactly match the cluster data. To examine whether this leads to a significant systematic uncertainty, we compare the predicted weight density to the observations of high redshift clusters beyond the extent of the cluster. \citetalias{hoekstra15} found that the contribution from cluster members and associated structures is less than 0.5\% for radii beyond 4 Mpc and we use this to define the areas in the \mn \ observations with mostly field galaxies. Due to the low redshift of \mn \ clusters there is very little area beyond 4 Mpc, so we check various outer radii for a more robust comparison.
Figure \ref{fig:blankcomp} shows the distribution of the difference between the weight density in the blank field prediction and in the cluster data. For all three outer radii $-$ 3.0 Mpc, 3.5 Mpc and 4.0 Mpc from left to right respectively $-$ the scatter is centered around zero. The coloured curves show the best fit Gaussian, which can be compared to the black curve, which shows the variation expected from the blank fields. We fixed the centre of the Gaussian on zero, but leaving the centre as a free parameter changes little in the fit.
The overall agreement between the scatter in the blank fields and the cluster outskirts is remarkably good. 
This again shows that our blank field prediction for the weight density of field galaxies is a reliable tool for the normalisation of the weight density in the cluster data.

\section{Results from cluster simulations}
\label{app:massbias}

\begin{figure*}
 \centering
 \includegraphics[width=17cm,height=8.5cm]{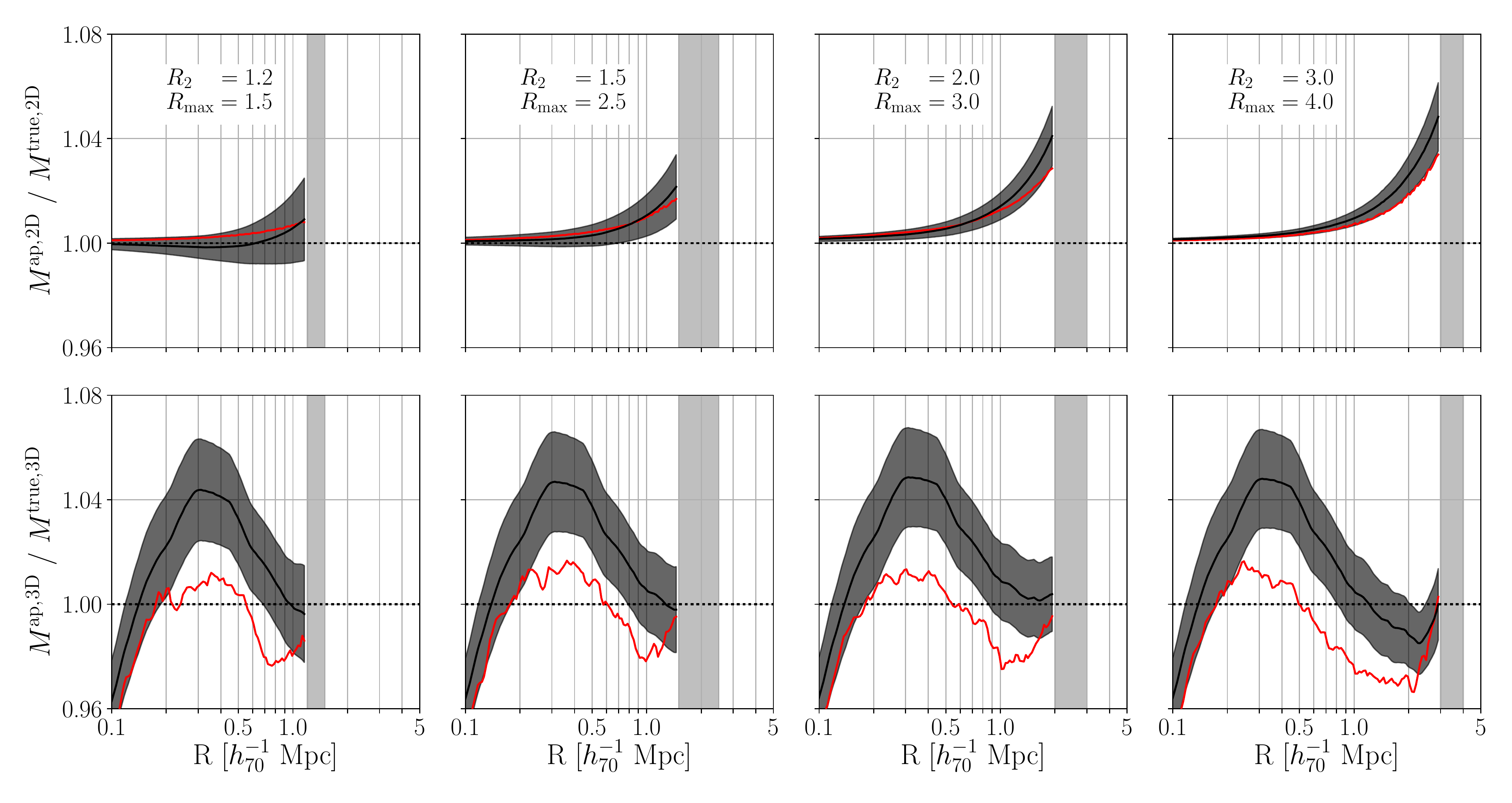}
 \caption{Mass bias for the projected aperture mass (top) and deprojected aperture mass (bottom) derived from the Hydrangea simulations for different choices of the annulus as a function of the distance to the true cluster centre. The vertical gray band shows the location of the annulus and is specified in the top panel. The average mass bias is shown in black with the black shaded region showing the uncertainty. The median mass bias for the 72 cluster profiles (24 clusters viewed from 3 angles) is shown in red, indicating that the distributions are highly skewed.}
 \label{fig:3d_mapbias}
\end{figure*}

\begin{table*}
\begin{tabular*}{1.0\textwidth}{ c c c c c c c|c c c c }
 \multicolumn{7}{ c|}{NFW mass} &  \multicolumn{4}{c }{Deprojected aperture mass}\\
radial fit range & \multicolumn{3}{ c }{$M_{200}$/$M^\mathrm{true}_{200}$} & \multicolumn{3}{ c|}{$M_{500}$/$M^\mathrm{true}_{500}$} &  \multicolumn{1}{c }{$R_2$, $R_\mathrm{max}$} & \multicolumn{3}{ c }{$M_{500}$/$M^\mathrm{true}_{500}$}  \\
Mpc & average & median & uncertainty &  average & median & uncertainty & Mpc  & average & median & uncertainty  \\
\hline
$0.5 \leq R \leq 1.5$ & 0.928   & 0.935  & 0.035  &  0.956  & 0.961  & 0.032  & 1.2, 1.5  & 1.004 & 0.970  & 0.021   \\
$0.5 \leq R \leq 2.0$ &  0.926   & 0.929  & 0.033  & 0.955   & 0.969  & 0.030  & 1.5, 2.5 & 0.977   & 0.960  & 0.021 \\
& & & & & & & 2.0, 3.0 & 0.975   & 0.965  & 0.019   \\
& & & & & & & 3.0, 4.0 & 0.981   & 0.945  & 0.021   \\
 \end{tabular*}
\caption{Statistics for the distribution of mass bias measured in the 24 Hydrangea clusters viewed along 3 axes for the NFW mass estimates (top) and the aperture mass estimates (bottom). The radii $R_{200}$ and $R_{500}$ are with respect to the critical density of the Universe. As we do not measure the aperture mass at $R_{200}$ for \mn, $M_{200}$ is not shown. Different physical sizes of the telescope field of view are explored to evaluate the mass bias for the nearby \mn \ clusters.  
} \label{tbl:massbias}
\end{table*}

Here we compute the systematic offsets in our two mass measurement methods: deprojected aperture masses and NFW fitting. A bias in the NFW mass comes from forcing the cluster to follow a specific spherically symmetric model which might not accurately describe its matter distribution.
In addition, our data does not allow us to fit for concentration freely, so we have to reduce the degrees of freedom of the NFW model by assuming a mass-concentration relation.
The aperture mass measurements suffer from the assumptions about the convergence in an annulus around the cluster and the mass distribution along the line of sight.
Moreover, our data puts constraints on the size and location of the annulus, as well as the fitting range for the NFW profile. The consequences of these choices can only be properly explored with mock data.

The Hydrangea cluster simulations \citep{bahe17, barnes17} provide an excellent test case for our data, as it realistically includes the effects of baryons, including AGN feedback, which can alter the cluster density profiles. The clusters sit at the centre of zoom-in regions extracted from a (3.2 co-moving Gpc)$^3$ volume, each extending to 10 $r_{200}$ from the cluster centre and simulated with the AGNdT9 variant of the EAGLE simulation model  \citep{schaye15, crain15}. For the 24 simulated clusters at $z=0$ in the Hydrangea simulations we created spherical mass profiles, by summing all the mass of all the species in the simulations inside the sphere, and circular mass profiles, by summing all the mass in cylinders of 16 Mpc  in length along each axis of the simulation box. The clusters are randomly oriented with respect to the box, so there is no preferential selection, such as of merging systems (although sometimes they do happen to lie along the line of sight). The cylinders are also long enough to capture the correlated structure associated to the cluster \citep{becker11}. The mass profiles are computed out to a radius of 5 Mpc. In total we have 72 cluster profiles spanning a mass range  $\mathrm{log}(M_{200}/M_\odot) = 14-15.4$, for $M_{200}$ with respect to the critical density of the universe, a similar range to \mn \ and CCCP.

From the 2D mass profiles we compute the shear $g$ using a dummy value for $\Sigma_\mathrm{crit}$, roughly the average for all \mn \ clusters, which is taken as known in the measurement. We do not investigate miscentring here and use the actual centres of the simulated clusters in our computations. We note that the central galaxies in the simulations are consistent with the center of the potential within the 1 kpc resolution of the simulations \citep{he19}. Using the BCG as the cluster centre, as we do for the observations, would therefore not change the results in this section. Masses are estimated in the same manner as was done for the data. First, the NFW profile is fit and $M^\mathrm{NFW}_{500}$ is calculated, and with the NFW mass the 2D aperture mass is computed and then deprojected to find $M^\mathrm{ap}_{500}$. In these simulations the Planck cosmology (flat Universe with $H_0 =67.77$ km/s/ Mpc and $\Omega_m =0.308$) was assumed and used in the mass measurements. The mass bias is then the measured mass within some radius divided by the true mass, which is calculated by summing the masses of all the particles inside that radius.

In the top panels of Figure \ref{fig:3d_mapbias} we show the 2D aperture mass bias as a function of aperture radius $R_1$ for various choices of the annulus, which is shown as the gray shaded area. The annuli shown are representative of the range in our data; the annulus in the left and right panels roughly correspond to our annulus choice at $z\approx0.05$ and $z\approx0.15$, respectively. For CCCP the rightmost panels are indicative. As expected if the annulus is placed far from the measurement radius the 2D aperture mass is almost unbiased, becoming less accurate and less precise as $R_1$ approaches $R_2$. Fortunately, even for the smallest choice of $R_2=1.2$ Mpc the biases is at most $\approx$1\% around 1 Mpc. In the bottom panels of Figure \ref{fig:3d_mapbias} we show the deprojected aperture mass profiles for the same choices of annulus. The deprojection introduces a much larger bias in the mean mass estimate up to 5\% depending on $R_1$. This shows that near the cluster core the NFW is a poor description, but also in regions farther out the deprojection  introduces percent level biases. 
We note that the median bias is very different from the mean bias, showing that the distribution of recovered mass is very asymmetric.
Around 1 Mpc both the mean and median bias are small, which is the range for our measured $R_{500}$ values.

We summarise the mass biases for the 3D mass estimates in Table \ref{tbl:massbias} for all radii which could be measured in the \mn \ data. We find good agreement with the literature. Our estimates of the bias are consistent with the findings of \citet{schrabback18} for their case of no miscentring for $M_{500}$. We see a slightly larger bias for $M_{200}$, although they are consistent within the errors. Our biases in $M_{200}$ and $M_{500}$ are similar to \citet{henson17} and \citet{becker11}, respectively. It is unclear why the aperture mass method performs much better for the 2D mass compared to the results of \citet{meneghetti10}, as we perform the same operations.

We repeated these exercises for a snapshot of the Hydrangea simulations at $z=0.3$, roughly the mean redshift for the CCCP clusters. We found very similar results as for the $z=0$ clusters, indicating a negligible evolution of the mass bias. We therefore assign the same bias correction to the CCCP clusters as to the \mn \ clusters. In contrast, the results of \citet{schrabback18} do show evolution in the mass bias between  $z=0.5$ and $z=1.0$. A possible explanation is that this evolution only happens at earlier times in the cluster formation, but our sample of simulated clusters is not large enough to draw a conclusion on this.

There is little difference between the average mass biases for large and small fields of view for the NFW mass estimates, although the medians show that the distribution is asymmetric. We use the average mass bias for a radial range out to 2 Mpc to correct all clusters in the data. The aperture masses also show small changes for different fields of view, all consistent within the uncertainties. As only a small fraction of the clusters has the smallest field of view, we correct all measured $M_{500}$ values in the data by dividing by 0.98.

\label{lastpage}
\end{document}